\documentclass[twocolumn]{aastex701}

\usepackage{amsmath}

\newcommand{\red}[1]{{#1}}

\usepackage{physics}

\begin{document}
 
\title{\red{A New Approach to Modeling Line Shapes with Quasi-H$_2^+$ Satellites in Stellar Atmospheres}}

\author[0000-0003-4052-2746]{Jackson R. White}
\affiliation{Department of Astronomy, University of Texas at Austin, Austin, TX 78712, USA}
\affiliation{X Computational Physics Division, Los Alamos National Laboratory, Los Alamos, NM 87545, USA}
\email[show]{jacksonw@lanl.gov}

\author[0000-0001-8748-5466]{Thomas A. Gomez}
\affiliation{Department of Astrophysical and Planetary Sciences, University of Colorado,  
 Boulder, CO 80305, USA}
\affiliation{Laboratory for Atmospheric and Space Physics, University of Colorado, Boulder, CO 80305, USA}
\affiliation{Department of Nuclear Engineering and Radiological Sciences, University of Michigan, Ann Arbor, MI 48109,
USA}
\email{thomas.gomez@colorado.edu}

\author[0000-0003-0473-379X]{Mark C. Zammit}
\affiliation{Theoretical Division, Los Alamos National Laboratory, Los Alamos, NM 87545, USA}
\email{mzammit@lanl.gov}
\author[0000-0002-6748-1748]{Michael H. Montgomery}
\affiliation{Department of Astronomy, University of Texas at Austin, Austin, TX 78712, USA}
\email{mikemon@astro.as.utexas.edu}

\author[0000-0002-1086-8685]{Bart H. Dunlap}
\affiliation{Department of Astronomy, University of Texas at Austin, Austin, TX 78712, USA}
\email{bhdunlap@utexas.edu}

\author[0000-0001-8816-236X]{Ivan Hubeny}
\affiliation{Department of Astronomy, University of Arizona, Tuscon, AZ, 85721, USA}
\email{ihubeny.astr@gmail.com}

\author[0000-0002-3951-9016]{Dmitry V. Fursa}
\affiliation{Department of Physics, Curtin University, Perth, Western Australia 6845, Australia}
\email{d.fursa@curtin.edu.au}

\author[0000-0001-7554-8044]{Igor Bray}
\affiliation{Department of Physics, Curtin University, Perth, Western Australia 6845, Australia}
\email{igor.bray@curtin.edu.au}

\author[0000-0003-0181-2521]{Don E. Winget}
\affiliation{Department of Astronomy, University of Texas at Austin, Austin, TX 78712, USA}
\email{dew@astro.as.utexas.edu}

\begin{abstract}
\red{Theoretical spectral line shapes describe the distribution of opacity due to bound electronic transitions in hot dense plasmas, and are used to fit emergent spectra from many astrophysical sources.} Deficiencies in line broadening theory have been proposed as a possible explanation for unresolved discrepancies between theoretical and observed spectra in white dwarf star atmospheres and laboratory experiments at white dwarf star photosphere conditions. One possible source of these discrepancies is the formation of quasi-molecules. Quasi-molecules are close (unbound) collisions between atoms, which broaden line shapes and create additional satellite lines. Quasi-molecules are challenging to implement into traditional line shape codes and have historically required a number of physical approximations beyond what is used in standard Stark broadening models. Here we present a new approach to calculating line shapes with quasi-molecular \red{H$_2^+$} resonances, using a novel multiple-basis method that considers both atomic and molecular states. We implement this approach into a simulation line shape code, present hydrogen Lyman-series line shapes with quasi-H$_2^+$ resonances, and demonstrate the impact our new line shapes have on hydrogen-atmosphere white dwarf star model spectra. We find that our new approach leads to broader quasi-molecular features that agree well with observed spectra in initial comparisons. 
\end{abstract}


\section{Introduction} \label{sec:intro}

White dwarf (WD) stars are the end stage of stellar evolution for almost all ($>\,$97\%) of the stars in our Milky Way galaxy \citep{VanHorn15}. A ubiquitous and long-lived stellar population, WD stars are uniquely applicable to many fields of astronomy. The oldest WDs help us date the age of the Milky Way, and the universe itself, through a process called white dwarf cosmochronology \citep{Winget87}. Space telescopes, including the Hubble Space Telescope and the James Webb Space Telescope, are flux calibrated using WDs \citep{Bohlin20,Gordon22}. Polluted WDs, which accrete the tidally disrupted fragments of former exoplanets, provide our only direct observations of exoplanet interiors \citep{Zuckerman07,Farihi13,Vanderbosch2020}. WD atmospheres can also be created in controlled laboratory environments \citep{Falcon13,Falcon15}, enabling robust validation of our astrophysical models by direct comparison with at-parameter experimental data and stellar observations.

Whether we leverage WDs as age markers, calibration sources, or  exoplanet interior spectrometers, their usefulness depends on our ability to determine precise stellar parameters, like their effective temperatures and surface gravities. The dominant techniques by which we estimate individual WD stellar parameters involve comparing theoretical synthetic spectra to astronomical observations \citep{Bergeron19}. 
These techniques include the `spectroscopic method', which involves fitting observed spectral lines to modeled spectral lines, and the `photometric method', which compares broadband photometric measurements to synthetic data.
Such fits are commonly done with optical wavelength observations, where hydrogen-atmosphere spectra are generally dominated by hydrogen Balmer-series spectral lines. Ultraviolet observations can also be used to perform these fits \citep{Sahu2023}, where hydrogen Lyman-series spectral lines tend to be the most prominent.
\red{Without accurate spectral line shape theory, these fits will provide inaccurate and inconsistent effective temperature, surface gravity, or metal abundance \citep{Allard2018,Blouin2019} estimates for WD stars, which can in turn affect how many other stellar characteristics are determined, like the mass \citep{Tremblay2009} or convective mixing efficiency \citep{tremblay2010}.}

\subsection{Stark Broadening in White Dwarf Stars}

Stark broadening theory describes the broadening of spectral lines by charged particle collisions, and is a necessary input for calculating synthetic WD spectra. 
Stark broadening is often the dominant line broadening mechanism in WD atmospheres \citep{Vidal1970,Tremblay2009,Saumon22}.
Unfortunately, a growing wealth of evidence suggests that modern Stark broadening theory is a significant source of error in WD spectral modeling. 
\citet{Fuchs2017} showed that individual Balmer lines within the same stellar atmosphere often give discrepant surface gravity and temperature estimates. 
\citet{GenestBeaulieu2019} identified systematic offsets in stellar parameters derived from fits to spectral lines and fits to broadband photometry. 
\citet{Schaeuble2019} found that density estimates of laboratory plasmas at WD photosphere conditions calculated from H$\gamma$ and H$\beta$ absorption lines differ by over 30\%. 
\citet{arseneau25} suggested that inadequate line shape modeling, particularly in the line wings, could be the cause of substantial biases in WD star radial velocity measurements.
Observations of polluted WDs \citep{Xu2019,Gansicke2012} have repeatedly noted large metal abundance differences when comparing optical and UV spectra. 
And \citet{Sahu2023} identified systematic discrepancies in both temperature and mass when comparing UV to optical spectra, with the mass discrepancies rising to almost 10\%.

These laboratory and observational inconsistencies have motivated the recent development of several new line shape models applicable to WDs \citep{Tremblay2020,Ferri21,Gomez2021,Cho2022,Stambulchik2022,Gomez2025}, but improvements are still needed for a wide range of conditions. Even the most modern line broadening codes are still reliant on a host of different approximations with often questionable validity \citep{Gomez2022}.
At optical wavelengths, hydrogen and helium laboratory data \citep{Schaeuble2019,Schaeuble22} remain poorly fit regardless of which line shape theory is used. And in the UV, the heavily saturated hydrogen Lyman series has proven particularly sensitive to line shape theory modifications \citep{Gomez2025}. In short, Stark broadening theory in WD stars ``remains unsatisfactory and a work in progress'' \citep{Saumon22}.

\subsection{Quasi-Molecular Resonances}
One complication to line shape modeling in WDs is the presence of quasi-molecules \citep{Allard1999}. A quasi-molecule is a transient close collision between atoms in a hot dense plasma that induces temporary molecular-like electronic structure, without creating a bound molecule. 
Arguably the simplest quasi-molecule is quasi-H$_2^+$, which forms during a collision between a neutral hydrogen atom and a free proton; it was also one of the first quasi-molecules to be observed and identified. Spectral features caused by quasi-H$_2^+$ and quasi-H$_2$ were originally noted in $\lambda$ Boötis stars \citep{Baschek1984} and hydrogen-atmosphere (DA) WD stars \citep{Koester1985,Nelan_Wegner1985}. The Unified Line Broadening Theory (ULBT) then successfully identified these lines as satellites associated with unbound H$_2^+$ and H$_2$ \citep{Allard_Koester_1992,holweger94}. These quasi-molecular resonances have also been observed in laboratory data \citep{Kielkopf1995}.

\begin{figure*}
    \centering
    \includegraphics[width=\textwidth]{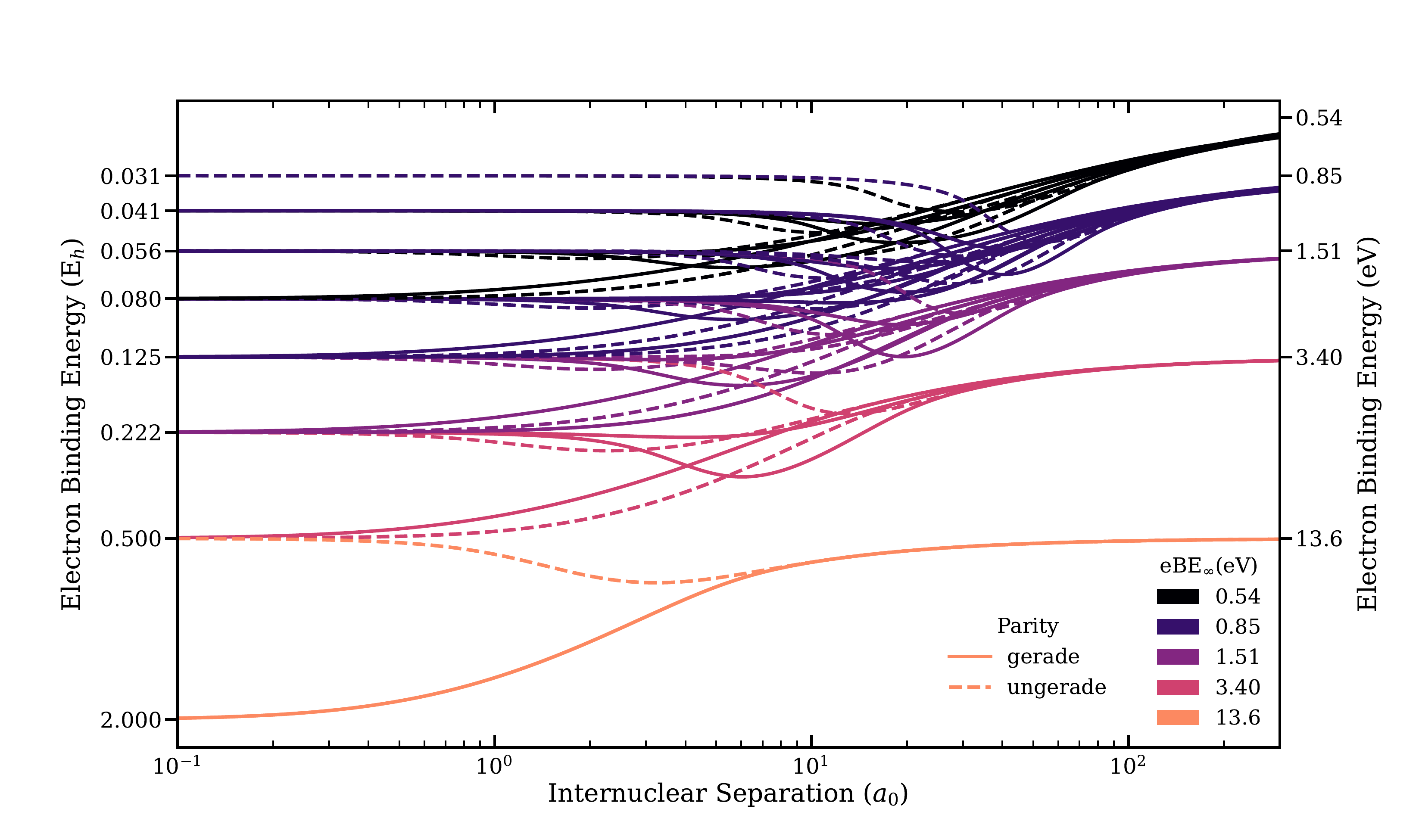}
    \caption{Electron binding energy (eBE) in an H$_2^+$ system as a function of internuclear separation \citep{Zammit2017,Zammit2018,Zammit2019}. Gerade states with even parity are shown with solid lines. Ungerade states with odd parity are shown with dashed lines. Line colors are set according to each state's electron binding energy at infinite internuclear separation.}
    \label{fig:h2p_curve}
\end{figure*}

The strongest quasi-molecular satellites in DAs are often the H$_2^+$ features associated with the hydrogen Lyman series.
Their inclusion in stellar atmosphere and spectral synthesis codes drastically improves the agreement between theory and observations \citep{Allard1998_1,Allard98_2,Allard2009}. In Fig.~\ref{fig:h2p_curve} we plot some excited state H$_2^+$ binding energy curves, as calculated in \citet{Zammit2017,Zammit2018,Zammit2019}, to demonstrate the non-trivial energy level splitting that occurs in quasi-H$_2^+$ at different internuclear spacings. Physically, quasi-molecular satellites can be understood as the result of local extrema in excited state molecular binding energy curves;
the local minima in the binding energy curves plotted in Fig.~\ref{fig:h2p_curve} generate a range of internuclear separations with similar transition energies, which cause quasi-H$_2^+$ systems in that range of internuclear separations to absorb photons at similar frequencies, thereby forming the satellites we see.

Quasi-molecules are challenging to model within line shape theory for the simple reason that most line shape theory is built around atomic states \citep{Gomez2022,Gomez24_PenetratingIons}. Stark broadening codes typically operate entirely in an atomic basis, have no way of accounting for the complex molecular structure evident in Fig.~\ref{fig:h2p_curve}, and neglect quasi-molecules entirely. Conversely, line broadening models that do incorporate quasi-molecules, like the ULBT, operate entirely in a molecular basis, where physical effects like screening and directional correlations are challenging to include.

In this paper we present a new approach; we have developed a multi-basis technique that alternates between an atomic basis and a molecular basis, and can be implemented in modern simulation-based Stark broadening codes to naturally incorporate quasi-molecular structure. 
\red{The present work is designed specifically for homonuclear diatomic singly charged quasi-molecules, like quasi-H$_2^+$.}
The rest of the paper is organized as follows. 
Sec.~\ref{sec:modeling} we discuss current techniques used to model quasi-molecular resonances. In Sec.~\ref{sec:Simulation_LS_Method} we briefly review standard line shape theory applicable to simulation-based line shape models. In Sec.~\ref{sec:New_LS_Method} we detail the theory of our novel multi-basis approach. 
In Sec.~\ref{sec:results} we present new hydrogen line shapes with quasi-molecular H$_2^+$ resonances as implemented in the \textsc{Xenomorph} code \citep{Cho2022}. In Sec.~\ref{sec:stellar_atmospheres_impact} we present model WD spectra calculated with our new lines. Finally, in Sec.~\ref{sec:conclusions} we present our conclusions. 
Atomic units are used throughout this paper unless otherwise stated, where $\hbar = e = m_e = 1$.

\section{Quasi-Molecular Models}\label{sec:modeling}

In this section we discuss two current approaches to calculating quasi-molecular resonances in line shapes; first the Unified Line Broadening Theory (ULBT) in \ref{sec:ULBT}, and then the Full-Coulomb (FC) method in \ref{sec:FC}. We then discuss some relevant physical line shape approximations in \ref{ULBT_approximations}.

\subsection{Unified Line Broadening Theory}
\label{sec:ULBT}
The ULBT \citep{Allard1999} is a semi-analytic line shape approach that successfully identified absorption features near the Ly$\alpha$ lines of DAs as satellites of two- and three-body quasi-molecules composed of hydrogen atoms and protons \citep{Allard1991}. 
It has since been applied to metal lines in brown dwarf stars \citep{Allard2003}, polluted WD stars \citep{Allard2014}, and to other hydrogen Lyman \citep{Allard2009} and Balmer \citep{Allard2022} lines.

The ULBT makes use of the Anderson-Talman approximation, which simplifies the line shape problem by representing the multi-perturber plasma with a single-perturber damping factor \citep{Royer1980}. 
The ULBT line shape $I_\alpha(\omega)$, for atomic transition $\alpha$ and photon frequency $\omega$, is given in \citet{Allard2000} as,
\begin{equation}
    I_\alpha(\omega) = \frac{1}{\pi}\,\mathfrak{R} \int^\infty_0 e^{n g_\alpha(s)} e^{-i\omega s} ds,
    \label{eq:ULBT_ls}
\end{equation}
with damping factor $g_\alpha(s)$,
\begin{multline}
    g_\alpha(s) = \frac{1}{\sum_{e,e'}^\alpha |d_{ee'}|^2} \sum_{e,e'}^{\alpha} \int_0^\infty 2\pi \rho \, d\rho \int^\infty_{-\infty} dx \, \Tilde{d}_{ee'}[R(0)] \\
     \bigg[ \mathrm{exp}\bigg( i\int^s_0 dt V_{e'e}[R(t)]\bigg) \Tilde{d}^*_{ee'}[R(s)]  - \Tilde{d}_{ee'}[R(0)]\bigg],
     \label{eq:g_eq}
\end{multline}
where $R(t)$ is the internuclear separation between the radiating atom (or ``radiator") and a single perturbing ion (or ``perturber") at time $t$, while $n$ is the density of perturbers, $s$ is an integration time, and $\mathfrak{R}$ denotes taking the real part. The perturber is assumed to move along a linear trajectory defined by impact parameter $\rho$, initial position $x$, and a single fixed velocity $v = \bar v = \sqrt{8kT/ \pi \mu} $, where $\mu$ is the reduced mass of the system. $V_{e'e}$ and $d_{ee'}$ respectively denote the $R$-dependent potential energy difference and \red{$R$-dependent} dipole transition moment between molecular state $e$ and $e'$. $\Tilde{d}_{ee'}$ denotes a dipole transition that is modulated by a Boltzmann factor. 
Sums over $\alpha$ are carried out over pairs of energy surfaces $(e, e')$  that approach the same infinite separation transition energy, $\omega_{ee'}\rightarrow \omega_{\alpha}$ as $R \rightarrow \infty$.

\red{Quasi-H$_2^+$ Lyman-series ULBT profiles relevant to WD stars have been published for Ly$\alpha$, Ly$\beta$, and Ly$\gamma$ \citep{Allard1998_1,Allard98_2,Allard2009}, making use of H$_2^+$ molecular data from \cite{Madsen71,Ramaker72, Allard2009}}. More recently, \citet{Pelisoli15} have published ULBT profiles, using the molecular data of \citet{Santos12}, for both the Lyman and Balmer series. 
\citet{LeboucherDalimier99} also developed a two-center semi-analytic molecular basis model, similar to the ULBT, however to our knowledge it has not been applied to lines relevant to WD atmospheres.

\subsection{Full-Coulomb Method}
\label{sec:FC}
The ion Full-Coulomb (FC) method \citep{Gomez24_PenetratingIons} is a recently developed approach that partially accounts for line broadening caused by close ion collisions. This method mirrors the electron broadening FC procedure, where multipole expansions used to calculate the electron broadening interaction potential are evaluated to infinite order following \citet{Waltz1984,Junkel2000}. When applied to ion collisions, the FC treatment generates resonances at approximately the correct wavelengths and with approximately the correct strength, without the need to precompute molecular data. 

Recent implementations of the FC method \citep{Gomez2021,Stambulchik2022} are however limited by their reliance on a single-center atomic basis, which is unable to capture physical effects like parity splitting. As a result there can be significant shifts in the location of the ion FC resonances \citep{Gomez24_PenetratingIons} compared to molecular predictions, so the method is currently advantageous only for high-Z quasi-molecules, where the necessary excited state molecular data is often unavailable or computationally prohibitive.

\subsection{Line Shape Approximations} \label{ULBT_approximations}

The ULBT is the current standard used in stellar atmosphere modeling. 
However it relies on a number of physical approximations that are generally not utilized in atomic spectral line shape models. 
Here we discuss a few of them and highlight when they could be of concern.

{\it Single velocity approximation:} Eq.~\eqref{eq:g_eq} is calculated for a single mean velocity $\bar v = \sqrt{8kT/\pi \mu}$ \citep{Allard1994} and does not include the full thermal distribution of perturber velocities. \citet{Allard1998_1} previously explored this approximation and found that it had a minor impact on the broadening in the line wings.
However, the conditions under which this approximation is viable have not been well established.

{\it Screening approximation:}
Spectral line shapes are sensitive to plasma particle correlations. 
These correlations are usually accounted for with Debye-screened Coulomb interactions, although some fully interacting line shape simulations have been performed \citep{Stambulchik07,Gigosos18}.
The ULBT, however, does not account for any plasma correlations, as it uses both non-interacting trajectories and unscreened molecular potentials. \red{This neglect of screening is justified during close collisions when the perturber separation is much smaller than a typical Debye length, $R \ll \lambda_{D}$, but not for perturbers that are farther away. Screening is therefore expected to be an important contribution to the line core, but unimportant for the far wings. }

{\it Separate broadening approximation:} Ion and electron broadening are calculated separately in the ULBT. Eq.~\eqref{eq:ULBT_ls} gives an ion-broadened line shape, which is later combined with an electron-broadened line shape \citep{Bohlin20}, using either a convolution or a simple addition. A simple addition is equivalent to a convolution in the far line wing, but distorts the total opacity in the line core. Convolving separate ion- and electron- broadened profiles neglects the correlations between the two broadening processes, and is not equivalent to including simultaneous broadening by both plasma particle species \citep{Stambulchik16}. \red{As with the previous screening approximation, this separate broadening approximation is expected to be well justified in the line wing, and poorly justified in the line core.}

{\it Field-angle approximation:} Stark broadened line shapes are sensitive to fluctuations in both the magnitude and direction of the perturbing electric field \citep{Calisti14,Demura2014,Stambulchik16}. 
However, the ULBT in Eq.~\eqref{eq:g_eq} only considers the magnitude of the internuclear separation $R = |\vec R|$. This neglects directional correlations, caused by the time evolving field angle. 

The impact of each of these approximations on quasi-molecular resonances is expected to be small individually. But the validity and limits of each approximation have remained largely untested. \red{In the following two sections we discuss our new change of basis procedure that allows us to at least partially remove each of the four approximations discussed above in Sec.~\ref{ULBT_approximations}.}

\section{Simulation Line Shape Theory} 
\label{sec:Simulation_LS_Method}
For a radiating atom in a hot dense plasma environment, the total power spectrum depends on the radiation frequency $\omega$, the speed of light $c$, and the normalized line shape function $I(\omega)$ \citep{Griem74},
\begin{equation}
    P(\omega) = \frac{4\omega^4}{3c^3} I(\omega).
\end{equation}
\red{The fundamental Stark-broadened line shape equation is given as the Fourier transform of the thermally averaged dipole autocorrelation function $C(t)$} \citep{Baranger58,Fano63},
\begin{equation}
    I(\omega) = \frac{1}{\pi}\,\mathfrak{R}\int_0^\infty \textnormal{d}t \; e^{\mathrm{i}\omega t} \; C(t),
    \label{eq:fls_1}
\end{equation}
\begin{equation}
    C(t) = \textnormal{Tr} \; \Big\{ \vec D \cdot \vec D(t) \rho \Big\},
    \label{eq:fls_2}
\end{equation}
where Tr\{\} denotes the trace operation, $\rho$ is the density matrix of the combined radiator + plasma system, $\vec D$ is the time independent dipole operator, and $\vec D(t)$ is the time evolved dipole operator expressed in the Heisenberg representation,
\begin{equation}
    \vec D(t) = U^\dag(t) \vec D U(t).
    \label{eq:fls_3}
\end{equation}
$U(t)$ denotes the time evolution operator, and is the solution of the time-dependent Schr\"{o}dinger equation,
\begin{equation}
    i \frac{d}{dt} U(t) = H(t) U(t),
    \label{eq:u_def}
\end{equation}
where $H(t)$ is a time-dependent Hamiltonian.
Formally, the trace in Eq.~\eqref{eq:fls_2} should be carried out over an infinite Hilbert space; this is not computationally feasible and a truncation of the states is required. \red{Typically, this truncated set of states is limited to a single charge state.}

\begin{figure}[]
    \centering
    \includegraphics[width=\columnwidth]{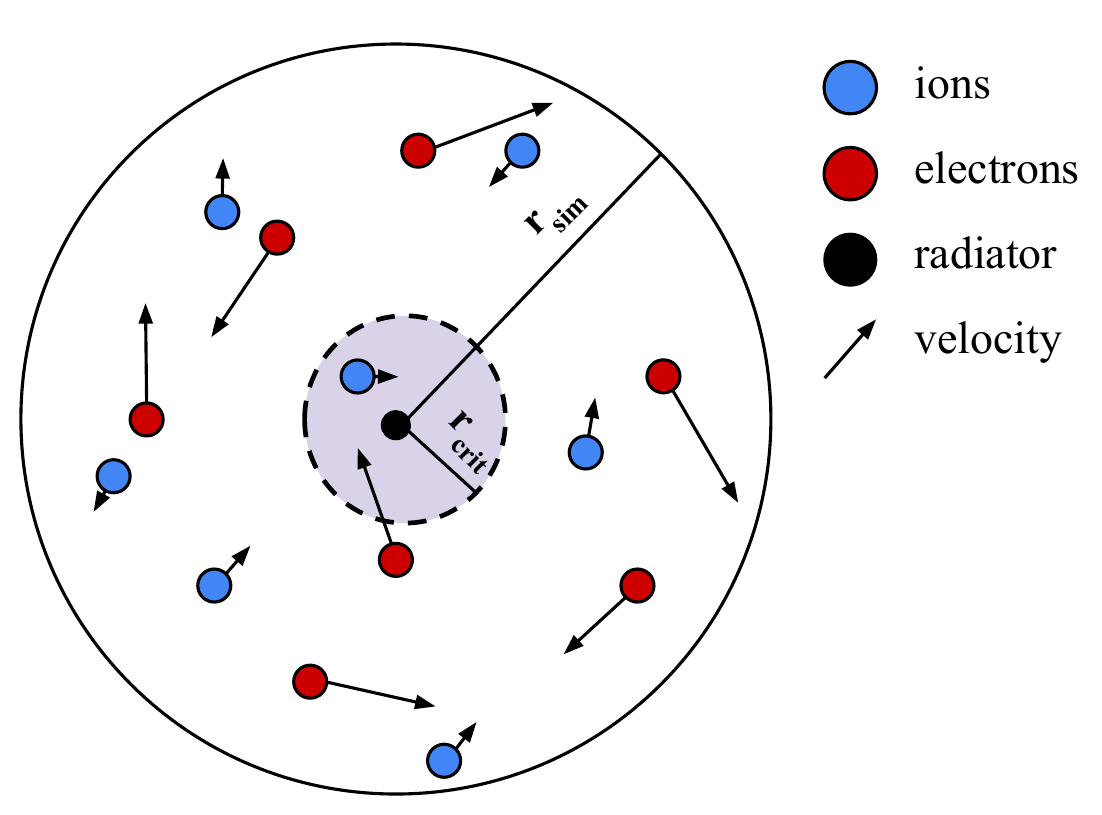}
    \caption{Two-dimensional cartoon of the simulation setup described in Sec.~\ref{cob_proc}. The simulation sphere is populated by electrons and ions moving on straight path trajectories, as described in \citet{Cho2022}. A critical radius $r_{\mathrm{crit}}$ is defined inside the simulation sphere, and we time evolve the system in a molecular basis when there is at least one ion within $r_{\mathrm{crit}}$. }
    \label{fig:xeno_cartoon}
\end{figure}

Simulation line shape codes \citep[e.g.][]{Cho2022,Tremblay2020,Stambulchik06,Gigosos87} use molecular dynamics simulations to calculate the $H(t)$ needed to solve Eqs.~\eqref{eq:fls_1} - \eqref{eq:u_def}. Most codes use the $\mu$-ion model, where the radiator is fixed at the center of a simulation while perturbing plasma particles move around it, as illustrated in Fig.~\ref{fig:xeno_cartoon}.
This approximation partially neglects the radiator motion, but is computationally advantageous and has been shown to agree well with more complete calculations \citep{Seidel82}. 
The Hamiltonian is taken to be that of an isolated atom $H_0$, plus a time-dependent potential $V(t)$ that accounts for the radiator-perturber interactions,
\begin{equation}
    H(t) = H_0 + V(t).
    \label{eq:ht_h0_vt}
\end{equation}
$V(t)$ can be calculated using a multipole expansion \red{\citep{Cowan1981}}. Taken to dipole order, the potential is a simple function of the time-dependent electric field $\vec \epsilon(t)$ felt by the radiator,
\begin{equation}
    V(t) = -\vec D \cdot \vec \epsilon(t),
    \label{eq:V_t_atomic}
\end{equation}
where $\vec D$ again denotes the atomic unperturbed dipole operator.
\red{Higher-order expansions are also possible \citep{Gomez16}. The time evolution operator of the radiator is then estimated by assuming that the Hamiltonian is constant across a given simulation time step $\Delta t$},
\begin{equation}
    U(t+\Delta t) \approx e^{-\mathrm{i}H(t)\Delta t} U(t) \label{time_evo_apx}.
\end{equation}

In the power spectrum method \citep{Stambulchik06}, the fundamental line shape equation is rewritten in terms of $\vec D(\omega)$, the Fourier transform of $\vec D(t)$,
\begin{equation}
    I(\omega) = \sum_{if} \rho_i \abs{\int_{-\infty}^{\infty} \textrm{d}t \, \vec D_{fi}(t) \, e^{\textrm{i} \omega t}}^2 = \sum_{if} \rho_i |\vec D_{fi}(\omega)|^2,
    \label{eq:fls_power}
\end{equation}
where $i$ and $f$ represent initial and final atomic states, respectively, and $\rho_i$ gives the population of state $i$. \red{This form is mathematically equivalent to a direct evaluation of the autocorrelation function, but is beneficial due to its rapid numerical convergence \citep{Rosato20}}.

\section{Multi Basis Line Shapes}
\label{sec:New_LS_Method}
A basis of electron configurations, or states, is needed to calculate a line shape profile. The most common choice is isolated single-center atomic states, where a perturbed radiator $\ket{r}$ can be represented as a superposition of isolated atomic states $\ket{a}$,
\begin{equation}
    \ket{r} \approx \sum_{a} c_a\ket{a},
    \label{eq:LinarComb}
\end{equation} 
where the expansion coefficients $c_a$ can be found by diagonalizing the $H(t)$. 
This choice is justified if the radiator is well separated from other ions, but breaks down if the radiator is not well isolated because the single-center basis cannot adequately model multi-ion (or `multi-center') electronic structure \citep{Gomez24_PenetratingIons,Zammit2013,Zammit2014}.
Taking the radial expectation value $\langle r_a \rangle$ as a characteristic wavefunction size for state $a$,
we expect single-center atomic states to be a good basis when $\langle r_a \rangle$ is much smaller than the nearest-neighbor ion separation $R$,
\begin{equation}
    \langle r_a \rangle \ll R.
    \label{eq:singleCenter_validity}
\end{equation}

The condition $\langle r_a \rangle \ll R$ naturally fails for highly excited configurations and in the limit of high densities, as bound electrons are gradually pressure ionized into the continuum.
When bound wavefunctions are comparable in size to the Wigner-Seitz radius $r_{ws}$,
\begin{equation}
    \langle r_a \rangle \approx R \approx r_{ws},
\end{equation}
spectral line shapes are observed to gradually broaden into the bound-free continuum \citep{Wiese72}.
Stellar atmosphere codes typically account for this modification of bound atomic structure with an ``occupation probability'' prescription \citep{Hummer1988} that gradually eliminates bound states as the density increases.

The condition $\langle r_a \rangle \ll R$ also fails in the case of unusually close atomic collisions, where $R$ is much smaller than usual due to random ion motion,
\begin{equation}
    \langle r_a \rangle \approx R \ll r_{ws}.
\end{equation}
This type of close collision is what we refer to as a quasi-molecule. 
Unlike the case of pressure ionization, electron wavefunctions of interest in diatomic quasi-molecules are assumed to overlap significantly with only one neighboring ion, rather than many neighboring ions.
In either case, a single-center atomic basis is inadequate for such line shape calculations.

\subsection{New Change of Basis Procedure} \label{cob_proc}

We have developed a multi-basis approach applicable to simulation line shape codes that alternates between a single-center atomic basis and a multi-center molecular basis, depending on the nearest-neighbor internuclear separation $R$. 
Here we describe this procedure and our implementation of it into the simulation line shape code \textsc{Xenomorph} \citep{Cho2022}.

We first define a critical radius $r_{\mathrm{crit}}$ around the radiator, as demonstrated in Fig.~\ref{fig:xeno_cartoon}.
(This choice is discussed further in Sec.~\ref{sec:crit_radius_choice})
When $R$ drops below $r_{\mathrm{crit}}$ at some critical time $t_c$ in the simulation, we transform our time evolution operator from an atomic basis $U^a$ into a molecular basis $U^m$,
\begin{equation}
        U^m_{\beta\alpha}(t_c) = \sum_{f,i} \bra{\beta}\ket{f}\bra{f} U^a(t_c) \ket{i}\bra{i}\ket{\alpha},
    \label{eq:initialCOB}
\end{equation}
where the sum is carried out over all pairs of initial and final atomic states $i$ and $f$ spanned by $U^a$, while $\alpha$ and $\beta$ denote molecular states spanned by $U^m$.
We then time evolve $U^m$ in the static-limit approximation, just as we do for $U^a$, following Eq.~\eqref{time_evo_apx},
\begin{equation}
    U^m(t+\Delta t) \approx e^{-\mathrm{i}H^m(t)\Delta t} U^m(t), \label{static_limit_mol}
\end{equation}
where $H^m(t)$ is a time-dependent \textit{molecular} Hamiltonian that includes the Hamiltonian of an isolated molecule $H^m_{0}(R)$ plus the interaction potential $V^m(t)$ between the isolated molecule and the surrounding plasma,
\begin{equation}
    H^m(t) = H^m_{0}(R) + V^m(t). \label{eq:hamil_mol_basis}
\end{equation}
Note that $H^m_{0}(R)$ here depends explicitly on the nearest-neighbor internuclear separation $R$, which is itself time dependent. Eq.~\eqref{eq:hamil_mol_basis} is simply a molecular version of Eq.~\eqref{eq:ht_h0_vt}.

In an atomic basis, $V(t)$ is calculated with a multipole expansion.
We adopt the same procedure for $V^m(t)$.
In this work we solve for $V^m(t)$ at dipole order, using the appropriate molecular dipole moments, $\vec D^m(R)$,
\begin{equation}
    V^m(t) \approx -\vec D^m(R) \cdot \vec \epsilon(t),
    \label{eq:V_t_mol}
\end{equation}
where $\vec \epsilon$(t) denotes the electric field felt by the isolated molecular system, analogous to Eq.~\eqref{eq:V_t_atomic}. Extensions to higher order multipoles are of course possible if higher order molecular multipole moments are calculated.

\begin{figure}[]
    \centering
    \includegraphics[width=1.05\columnwidth]{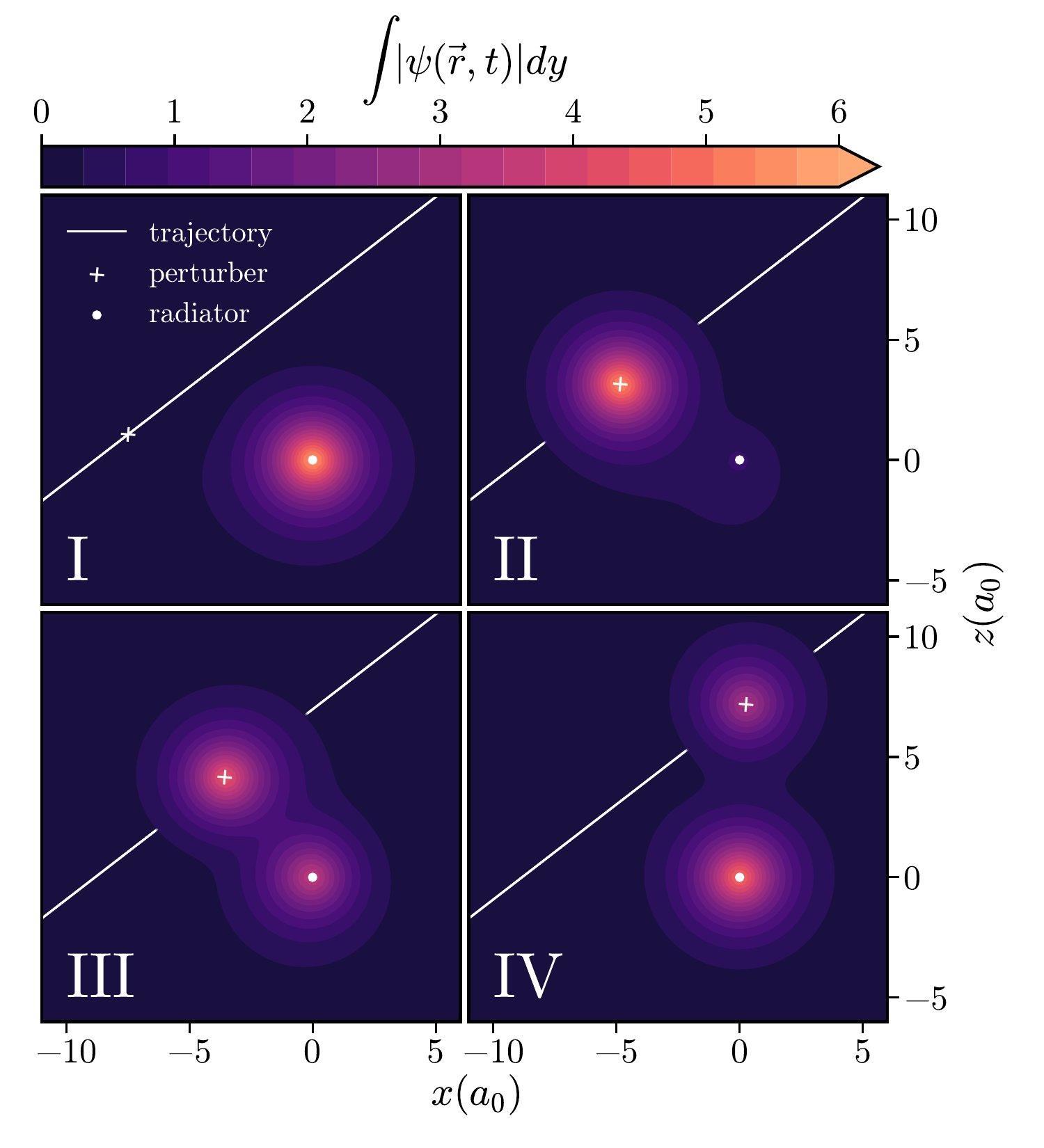}
    \caption{Electron wave function snapshots at four different times (I corresponds to the earliest timestep, IV to the latest timestep) during a quasi-H$_2^+$ close collision. The bound electron is initially assumed to be in an atomic 1$s$ state around the radiator (marked by a white dot and fixed at the origin). During the close collision, the electron is free to move between the `radiator' proton and the `perturber' proton (marked by a white cross and moving along a straight path trajectory). Asymmetries in the wavefunction are present due to perturbations from other distant plasma particles in the simulation. }
    \label{fig:charge_ex}
\end{figure}

One benefit of time evolving our system in a molecular basis is that we naturally account for charge exchange. In Fig.~\ref{fig:charge_ex} we present snapshots of the time evolved wavefunction,
\begin{equation}
    \psi(t) = U(t) \psi(0),
\end{equation}
where $U(t)$ is taken from a \textsc{Xenomorph} simulation with our multi-basis approach implemented for quasi-H$_2^+$ collisions in a hydrogen plasma. $\psi(0)$ is assumed to be a 1$s$ state for the purpose of the figure.

At the end of a close collision, when $R$ rises above $r_{\mathrm{crit}}$ at some time $t_e$, we move back into an atomic basis, 
\begin{equation}
        U^a_{fi}(t_e) = \sum_{\beta,\alpha}\bra{f}\ket{\beta}\bra{\beta}U^m(t_e) \ket{\alpha}\bra{\alpha}\ket{i},
    \label{eq:secondCOB}
\end{equation}
and we carry on with the simulation in an atomic basis until the next close collision.

\subsection{Unitarity and Inter-Atomic Transitions}
\label{sec:unitarity}
Here we discuss how we preserve the unitarity of the time evolution operator $U(t)$ throughout our change of basis procedure.
Mathematically, $U(t)$ is unitary if
\begin{equation}
    U^\dag(t) U(t) = I,
\end{equation}
where $I$ is the identity matrix.
Physically, the unitarity of $U(t)$ implies conservation of electron number in our simulations.
A general requirement of simulation line shape codes is therefore to preserve unitarity.
$U(t)$ is initially assumed to be the identity matrix,
\begin{equation}
    U(0) = I,
    \label{eq:U_init_condit}
\end{equation}
and is then time evolved through Eqs.~\eqref{time_evo_apx} or \eqref{static_limit_mol}, using the appropriate time-dependent Hamiltonian. As our Hamiltonian is Hermitian, we preserve the unitarity of $U(t)$ to machine precision while time evolving it in \textsc{Xenomorph}.

The challenge we face is preserving unitarity across a change of basis, i.e.~when carrying out Eqs.~\eqref{eq:initialCOB} and~\eqref{eq:secondCOB}. 
Our change of basis procedure only preserves unitarity if we use equivalent atomic and molecular bases, where every state in the molecular basis $\ket{m}$ is a linear combination of atomic states $\ket{a}$ and every atomic state is a linear combination of molecular states, i.e.,
\begin{equation}
    \ket{m} = \sum_a \ket{a}\bra{a}\ket{m},
    \label{eq:mstate}
\end{equation}   
\begin{equation}
    \ket{a} = \sum_m \ket{m}\bra{m}\ket{a}.
    \label{eq:astate}
\end{equation}
If we do not use equivalent bases then the time evolution operator $U(t)$ will not be fully represented across a change of basis, and we will lose unitarity.
    
We impose two requirements to ensure that the bases are equivalent.
First, we assume infinite separation overlap integrals,
\begin{equation}
    \braket{a}{m}_{R = r_{\mathrm{crit}}} \approx \braket{a}{m}_{R=\infty}.
    \label{eq:inf_sep_overlaps}
\end{equation} 
This approximation is necessary for Eqs.~\eqref{eq:mstate} and~\eqref{eq:astate} to be finite sums. 
And this approximation is well justified if $r_{\mathrm{crit}}$ is large compared to the atomic wavefunction size.
Second, we extend the atomic basis to include both radiator \textit{and} perturber atomic states for the nearest-neighbor ion.
This doubles the size of the atomic basis. 
Such a dual-center atomic basis is not orthogonal for any finite internuclear separation, however we already assumed infinite internuclear separations for the purpose of calculating overlap integrals in Eq.~\eqref{eq:inf_sep_overlaps}. In that infinite separation limit the radiator and perturber atomic wavefunctions have no overlap and the basis is orthogonal.

The extended atomic basis matrices can be written as block matrices, with four distinct partitions, corresponding to whether the initial and final states are radiator or perturber states.
The time evolution operator, for example, can be written as
\begin{equation}U^a(t) = \begin{bmatrix}
U_{rr}(t) & U_{rp}(t) \\
U_{pr}(t) & U_{pp}(t) \\
\end{bmatrix},
\end{equation}
where $U_{rr}$ denotes all the matrix elements with radiator initial states and radiator final states (i.e.~the normal $U^a(t)$ matrix we usually calculate in the $\mu$-ion model). To be consistent with the initialization (Eq.~\ref{eq:U_init_condit}), the rest of the time evolution operator is assumed to be the identity matrix when the first close collision begins at time $t_{c1}$, 
\begin{equation}U^a(t_{c1}) \approx \begin{bmatrix}
U_{rr}(t_{c1}) & 0 \\
0 & I & \\
\end{bmatrix}.
\end{equation}
By writing the atomic time evolution operator in this extended basis, and by assuming infinite-separation overlap integrals, the atomic and molecular bases are now equivalent.
Eqs.~\eqref{eq:mstate} and~\eqref{eq:astate} now both hold, and we can preserve the unitarity of $U(t)$ to machine precision even when performing a change of basis. 

We note that $U^a$ is no longer block diagonal after a close collision ends (Eq.~\ref{eq:secondCOB}). 
The off-diagonal blocks of this post-collision $U^a(t)$ are non-zero due to charge exchange, as is evident in Fig.~\ref{fig:charge_ex}. 
We discuss some complications related to these off-diagonal blocks in appendices \ref{app:unitarity} and \ref{app:preserving_mu}.

One natural consequence of these off-diagonal blocks in $U^a(t)$ is that the time evolved atomic dipole matrix $D^a(t)$ also includes off-diagonal blocks,
\begin{equation}
    D^a(t) = \begin{bmatrix}
D_{rr}(t) & D_{rp}(t) \\
D_{pr}(t) & D_{pp}(t)
\end{bmatrix}.
\label{eq:Dt_block_form}
\end{equation}
In this work we refer to the transitions described by these off-diagonal blocks as ``inter-atomic'' transitions, as they are labeled with initial and final states associated with different atoms.

\subsection{Dipole Moments} \label{sec:line_definition}

During a close collision we calculate the time evolved molecular dipole moment $\vec D^m(t)$ using a molecular version of Eq.~\eqref{eq:fls_3},
\begin{equation}
    \vec D^m(t) = U^{m}(t)^\dag \vec{D}^{{\Tilde{m}}}(R) U^{m}(t).
\end{equation}
Following the ULBT \citep{Allard2004} we use a modulated molecular dipole moment $\vec{D}^{{\Tilde{m}}}(R)$,
\begin{equation}
    \vec D^{\tilde{m}}_{\beta \alpha}(R) = \vec D^{m}_{\beta \alpha}(R) e^{-\Delta V_{\alpha}(R) / {2kT}},
\end{equation}
where $\vec D^{m}_{\beta \alpha}(R)$ denotes the unperturbed dipole moment from initial molecular state $\alpha$ to final molecular state $\beta$ at internuclear separation $R$.
This unperturbed molecular dipole moment is modulated by a Boltzmann factor based on the change in potential energy of initial molecular state $\alpha$,
\begin{equation}
    \Delta V_\alpha(R) = V_{\alpha}(R) - V_{\alpha}(\infty),
\end{equation}
\red{which helps account for deviations from the perturbing ion's straight path trajectory.
This modulation can also be interpreted as correcting for the uniform perturber densities otherwise assumed by both the ULBT and our simulations.} 
Note that $V_{\alpha}(R)$ is the potential energy of the isolated molecule in state $\alpha$ with internuclear separation $R$.
$V_{\alpha}(R)$ is not the interaction potential given in Eqs.~\eqref{eq:V_t_mol} or~\eqref{eq:V_t_atomic}.

To calculate the final line shape profile (Eq.~\ref{eq:fls_power}), the time evolved dipole moment $\vec D(t)$ must be defined in a consistent basis across the entire simulation time. It is currently more convenient in stellar atmosphere codes to define `atomic' line shapes where all the bound-bound opacity is labeled by an initial and final atomic state. 
We therefore perform a change of basis to move $D^m(t)$ into an atomic basis after it has been calculated,
\begin{equation}
    D^a_{fi}(t) = \sum_{\beta,\alpha} \bra{f}\ket{\beta}\bra{\beta} D^m(t) \ket{\alpha}\bra{\alpha}\ket{i},
    \label{eq:Dt_cob}
\end{equation}
using the same procedure described in Sec.~\ref{sec:unitarity}.
The overlaps are again calculated at infinite internuclear separations, and the atomic basis again spans both radiator and perturber states. 

Unlike our $U(t)$ change of basis, which is only done at the critical radius, Eq.~\eqref{eq:Dt_cob} must be performed throughout the close collision. 
Our justification for using infinite separation overlaps is therefore no longer valid as $R$ can be arbitrarily small. 
However, we still use the infinite separation overlap integrals for a practical reason --- they make the bookkeeping tractable. 
A molecular state, at small $R$, will overlap with a number (formally an infinite number) of different atomic states. 
$R$-dependent overlaps in Eq.~\eqref{eq:Dt_cob} would then split the quasi-molecular opacity across countless different atomic line profiles. 
This is computationally impractical, so we retain the infinite separation overlaps for the time being. 
We note that this approximation is similar to the ULBT's choice to label molecular transitions according to their infinite separation energies.

\subsection{Molecular Data}

We must precompute molecular data over a range of internuclear spacings in order to carry out these line shape calculations in a molecular basis.
Potential energy curves, multipole moments, and wavefunction overlaps are all necessary inputs.
In this work where only quasi-H$_2^+$ collisions are considered, we have relied on the H$_2^+$ data from \citet{Zammit2017,Zammit2018,Zammit2019} as demonstrated in Fig.~\ref{fig:h2p_curve}, as well as the 2D Schr\"{o}dinger solver outlined in \cite{Gomez18a}. All molecular data for this work was calculated in the Born-Oppenheimer approximation. 

\section{Results} \label{sec:results}

Here we present quasi-H$_2^+$ line shapes calculated with the change of basis procedure outlined in the preceding section, having implemented it in the simulation line shape code \textsc{Xenomorph} \citep{Cho2022}.
We show Ly$\alpha$ and Ly$\beta$ profiles, at conditions relevant to DA WD photospheres, including results for both the far line wings and the line cores. The Lyman series in WDs is heavily saturated, so their stellar spectra are generally more sensitive to the far line wings. However, accurate line cores are still of interest \citep{Santos12,Pelisoli15}, particularly because the Balmer lines in WD stars are unsaturated.

In Sec.~\ref{sec:ULBT_comp} we compare our line shapes to the ULBT, and we demonstrate that we can recover the ULBT profile with a simplified implementation of our method.
In Secs.~\ref{sec:dir_correlations} through \ref{sec:crit_radius_choice} we test a number of different line shape approximations to understand how impactful they are on our profiles.
Finally, in Sec.~\ref{sec:inter_atomic_results}, we show how the ``inter-atomic'' transitions discussed in Sec.~\ref{sec:line_definition} contribute to the line shape.
\red{Excepting the density and temperature comparisons in Figs.~\ref{fig:density_convergence} and \ref{fig:temp_dependence}, our line shape calculations in this section are performed at $\textrm{T}=10,000~\textrm{K}$ and $n_e = 10^{17} $~cm$^{-3}$}.

\begin{figure}
    \centering
    \includegraphics[width=\columnwidth]{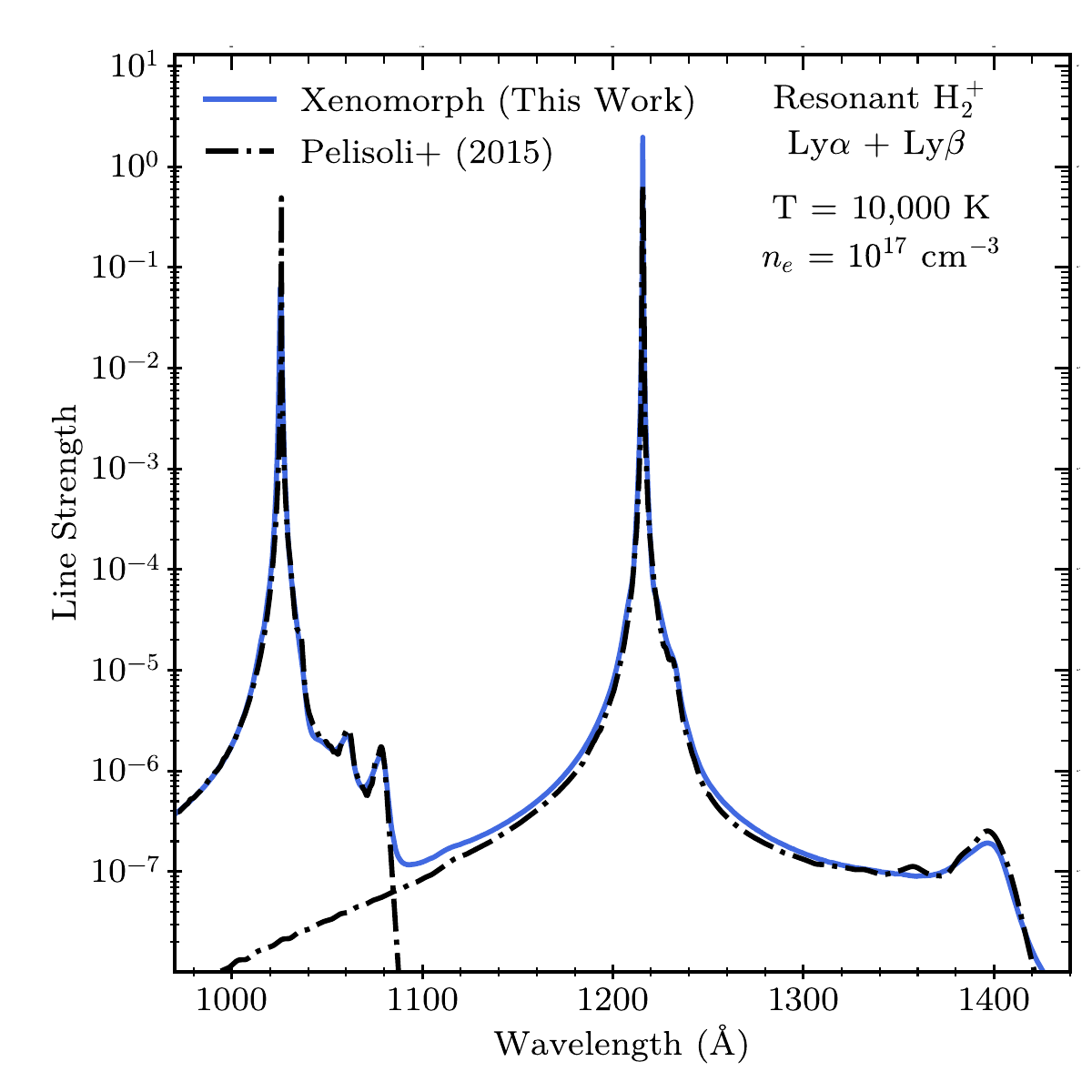}
    \caption{ Quasi-H$^+_2$ Ly$\alpha$ and Ly$\beta$ line shape profiles calculated with the ULBT (dot-dash) with our new multi-basis approach implemented in the \textsc{Xenomorph} code (solid).}
    \label{fig:pelisoli_comp_wide}
\end{figure}

\begin{figure*}
    \centering
    \includegraphics[width=\textwidth]{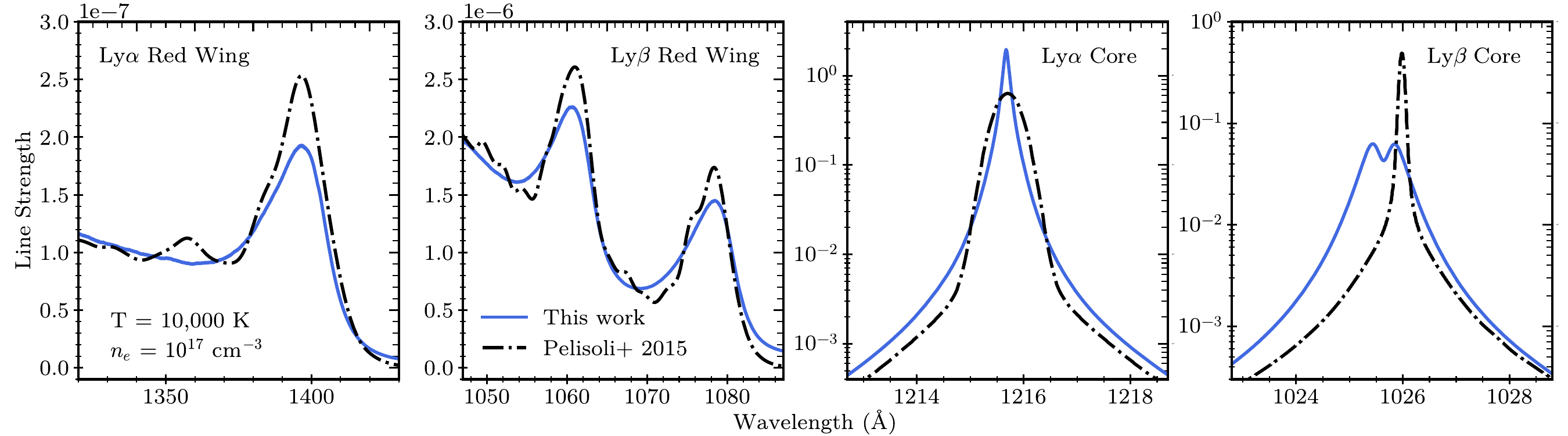}
    \caption{Four different sections of the ion-only line shape comparison presented in Fig.~\ref{fig:pelisoli_comp_wide}, highlighting the red wings and the cores of both Ly$\alpha$ and Ly$\beta$, with H$_2^+$ resonances present.}
    \label{fig:pelisoli_panel_comp}
\end{figure*}

\begin{figure}
    \centering
    \includegraphics[width=\columnwidth]{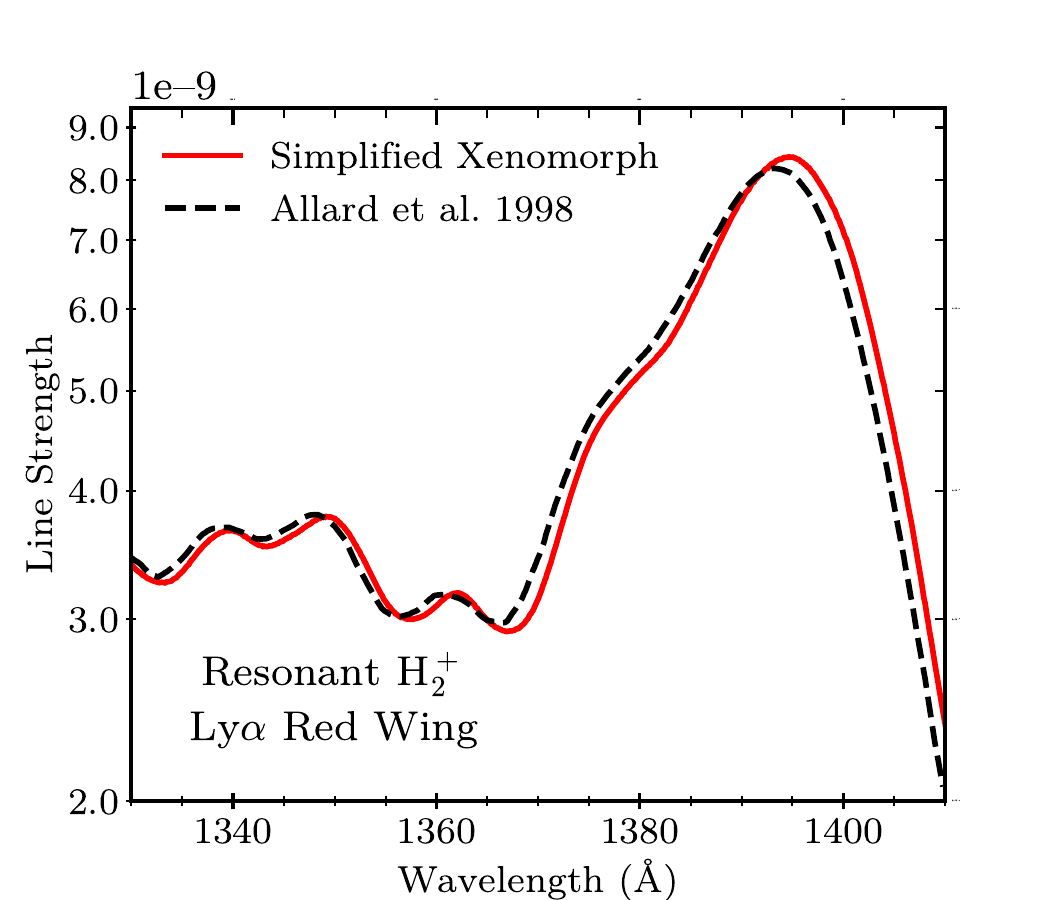}
    \caption{ Comparison of quasi-H$_2^+$ Ly$\alpha$ red wings between the ULBT (dash) from \citet{Allard1998_1}, and a simplified version of the \textsc{Xenomorph} code (solid), where we have calculated a single-site line with the single-velocity approximation, neglected electrons, removed screening, used the Boltzmann weighting as described in \citet{Allard1998_1}, and normalized the profile arbitrarily to match the original figure.}
    \label{fig:allard_rec}
\end{figure}

\subsection{Comparison to the ULBT} \label{sec:ULBT_comp}

Figs.~\ref{fig:pelisoli_comp_wide} and \ref{fig:pelisoli_panel_comp} compare our new Ly$\alpha$ and Ly$\beta$ lines against those published in \citet{Pelisoli15}. These profiles are `ion-only', meaning ion broadening is included while electron broadening is neglected. Each profile is area normalized and weighted by the appropriate atomic oscillator strength. Comparisons between the quasi-H$_2^+$ features and the line cores are shown in greater detail in Fig.~\ref{fig:pelisoli_panel_comp}, while the total profiles are shown in Fig.~\ref{fig:pelisoli_comp_wide}. 

The overall agreement between methods is good; the profiles are generally within a factor of two of each other, even though Fig.~\ref{fig:pelisoli_comp_wide} spans over eight orders of magnitude in line strength. 
A few key differences are however evident. While the ULBT profiles both have strong central cores, our Ly$\beta$ line is double peaked and missing its central component, as is expected for a hydrogen Ly$\beta$ line at these conditions \citep{Ferri14}. Our molecular resonances also appear to be broader, and much of the small-scale structure present in the ULBT wings is washed out (e.g., we do not find a secondary line peak at $\sim$1360Å).

Some of the line shape differences can be explained by the fact that we have removed a number of approximations that are typically present in the ULBT. For example the smoother molecular resonances \red{(which agree well with the observational data shown later in Ch.~\ref{sec:stellar_atmospheres_impact})} appear to largely be a result of incorporating a thermal velocity distribution, as discussed in Sec.~\ref{sec:sinlge_vel_approx}. However, we caution that the approximations explored in this paper are far from exhaustive, and additional approximations remain in both methods that can contribute to the discrepancies in Figs.~\ref{fig:pelisoli_comp_wide} and~\ref{fig:pelisoli_panel_comp}.

In Fig.~\ref{fig:allard_rec} we demonstrate that we can successfully replicate the structure of the ULBT quasi-H$_2^+$ Ly$\alpha$ red wing from \citet{Allard1998_1}, by artificially introducing the approximations that the ULBT relies on.

\subsection{Directional Correlations} \label{sec:dir_correlations}
Spectral lines are sensitive to both ``vibrational'' broadening, caused by fluctuations in the magnitude of the local electric field, as well as ``rotational'' broadening, caused by changes in the direction of the electric field \citep{Calisti14,Demura2014,Stambulchik16}. We will generally use the terminology ``directional correlations'' instead of rotational broadening, to distinguish this broadening from bulk Doppler broadening or rotational molecular state effects. In the static limit, the directional correlations can be neglected by averaging over the field angles and aligning the perturbations with the $z$-axis \citep{Iglesias13}. However, it has long been shown that directional correlations modulate the lines when ion-motion is included \citep{Demura77}.  

\begin{figure}
    \centering
    \includegraphics[width=\columnwidth]{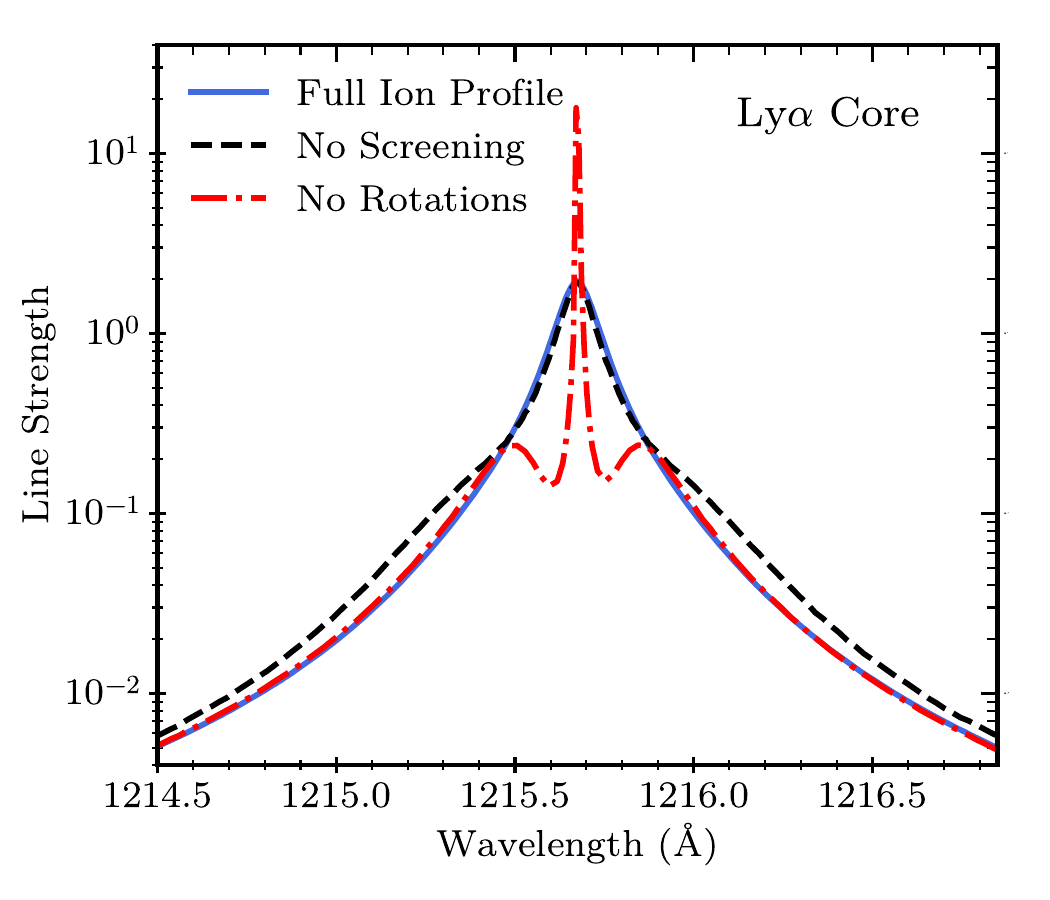}
    \caption{Three ion-only quasi-H$_2^+$ Ly$\alpha$ \textsc{Xenomorph} line shapes calculated at $\textrm{T}=10,000~\textrm{K}$ and $n_e = 10^{17} $~cm$^{-3}$, and zoomed in to show the line core. The red dot-dash line includes no directional correlations, or no ``rotational broadening''. The black dashed line includes no screening. The blue line is the full profile with no extra approximations artificially added in.}
    \label{fig:screen_rot_comp}
\end{figure}

We automatically account for directional correlations through our calculation of the perturbing potential $V(t)$ --- note that Eqs.~\eqref{eq:V_t_atomic} and~\eqref{eq:V_t_mol} consider the direction of the perturbing electric field, not just the magnitude. 
Our code is the first to our knowledge to incorporate any directional correlations into a quasi-molecular resonance calculation.
As the far wings are in the static limit, directional correlations primarily affect the line shape core, as is demonstrated in Fig.~\ref{fig:screen_rot_comp}.

We caution that we are still neglecting some of the directional correlations. $H^m_0(R)$ in Eq.~\eqref{eq:hamil_mol_basis} depends only on the magnitude of the internuclear separation, $R=|\vec R|$, and not on the direction.

\subsection{Screening} \label{sec:screening}
We account for particle correlations in our simulations with a Debye-screened Coulomb interaction. The perturbing electric field $\vec \epsilon$ is calculated as
\begin{equation}
    \vec \epsilon = \sum_p \vec \epsilon_p = \sum_p \frac{q_p}{r_p^2} \left( 1+\frac{r_p}{\lambda_D}\right) e^{-r_p/\lambda_D} \hat{r}_p,
\end{equation}
where $q_p$ is the charge of the $p$th perturber, $r_p$ is the distance between the radiator and perturber $p$, and $\hat{r}_p$ gives the direction of the perturber relative to the radiator. 
We assume that both ion and electron perturbers are screened only by electrons, so their Debye lengths are equivalent, $\lambda_{D}$ = $\lambda_{D,i}$ = $\lambda_{D,e}$. As with the directional correlations, screening has a moderate effect on the line core, which is demonstrated in Fig.~\ref{fig:screen_rot_comp}.

The nearest-neighbor ion is excluded from the above summation during a close collision, as its perturbation is accounted for in the (unscreened) molecular Hamiltonian $H_0^m(R)$.
We can safely ignore screening in $H_0^m(R)$ as long as the critical radius is small compared to the screening length. For reference, at T = 10,000 K, and $n_e = 10^{17}$ cm$^{-3}$, the screening length is $\lambda_{D} \approx 412 \, a_0$ and the $1400$Å feature forms due to quasi-molecules at $ R\approx 10 \, a_0$ (Fig.~\ref{fig:h2p_curve}).

\subsection{Mean Velocity Approximation}\label{sec:sinlge_vel_approx}

\begin{figure}
    \centering
    \includegraphics[width=\linewidth]{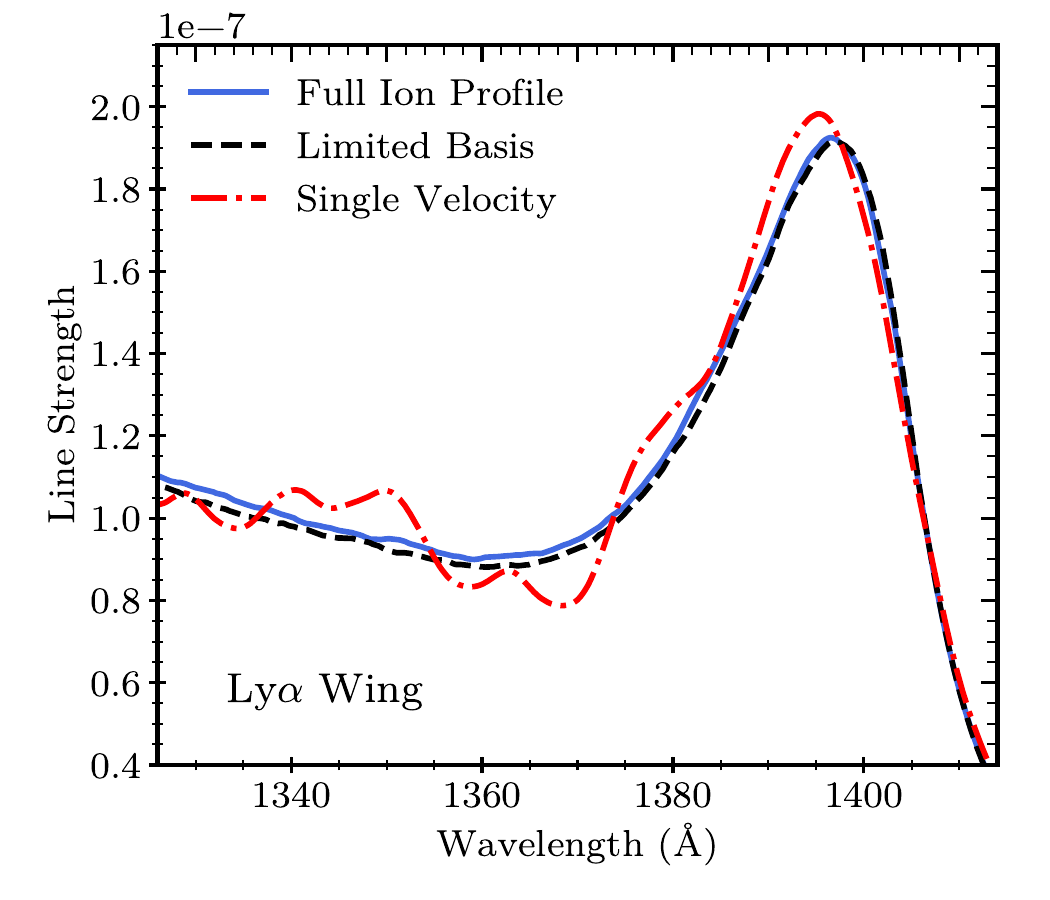}
    \caption{{\sc{Xenomorph}} ion-only quasi-H$_2^+$ Ly$\alpha$ line shapes at $\textrm{T}=10,000~\textrm{K}$ and $n_e = 10^{17} $~cm$^{-3}$. The red dot-dash line was calculated with the mean velocity approximation (Sec.~\ref{sec:sinlge_vel_approx}), as opposed to a Maxwellian distribution function. The black dashed line was calculated using a limited basis (Only n=1 and n=2 states are included, Sec.~\ref{sec:ext_basis}). The blue line is the full profile (with $n$=1, 2, and 3 states included in the basis). }
    \label{fig:thermal_basis}
\end{figure}

In the mean velocity approximation the speed of each perturbing ion is assumed to be the same mean thermal velocity $v = \bar v = \sqrt{8kT/\pi \mu}$ \citep{Allard1998_1}. \red{We remove this approximation in our present calculations by always populating our simulation sphere with ion and electron velocities randomly drawn to match the expected thermal distribution, as detailed in \citet{Cho2022}.} In Fig.~\ref{fig:thermal_basis} we test the validity of the mean velocity approximation by artificially introducing it into our model, and find that the mean velocity approximation introduces substantial small scale structure in the quasi-H$_2^+$ features.

\subsection{Expanded Basis} \label{sec:ext_basis}

We make use of an expanded atomic basis set in our calculations. An expanded basis is one that includes atomic states beyond the principal quantum numbers of the initial and final state for some atomic transition. This is motivated by the mixing between states, which adds to the broadening of the system \citep{Kilcrease93}. In Fig.~\ref{fig:thermal_basis} we show the difference between using a limited basis set (\red{bound H states with an electron in $n$=1 or $n$=2)} and an expanded basis set (\red{bound H states with an electron in $n$=1, 2, or 3)}. We find that the expanded basis only marginally modifies the line strength in the far wing.

\subsection{Simultaneous Ion + Electron Broadening}
Of practical concern is the combination of ion and electron broadening. In the ULBT, ion-only line shapes are calculated and must later be combined with an electron-only line shape. The adopted procedure in most stellar atmosphere codes, as in \citet{Hubeny95} and \citet{Koester1985}, is to convolve the ion-only line shape with an electron-only line shape. In practice the convolution is replaced by a simple addition, which is equivalent to a convolution in the far wing.
However, a convolution of an electron-broadened line and an ion-broadened line is not equivalent to modeling the simultaneous two-component plasma \citep{Stambulchik16}, as electron and ion broadening are not independent processes.
The simulation sphere in the present calculations is populated by ions and electrons that both contribute to the time evolution of the radiator, so we capture these simultaneous broadening correlations and eliminate the need to add in a separate electron-only profile.

\begin{figure}
    \centering
    \includegraphics[width=\columnwidth]{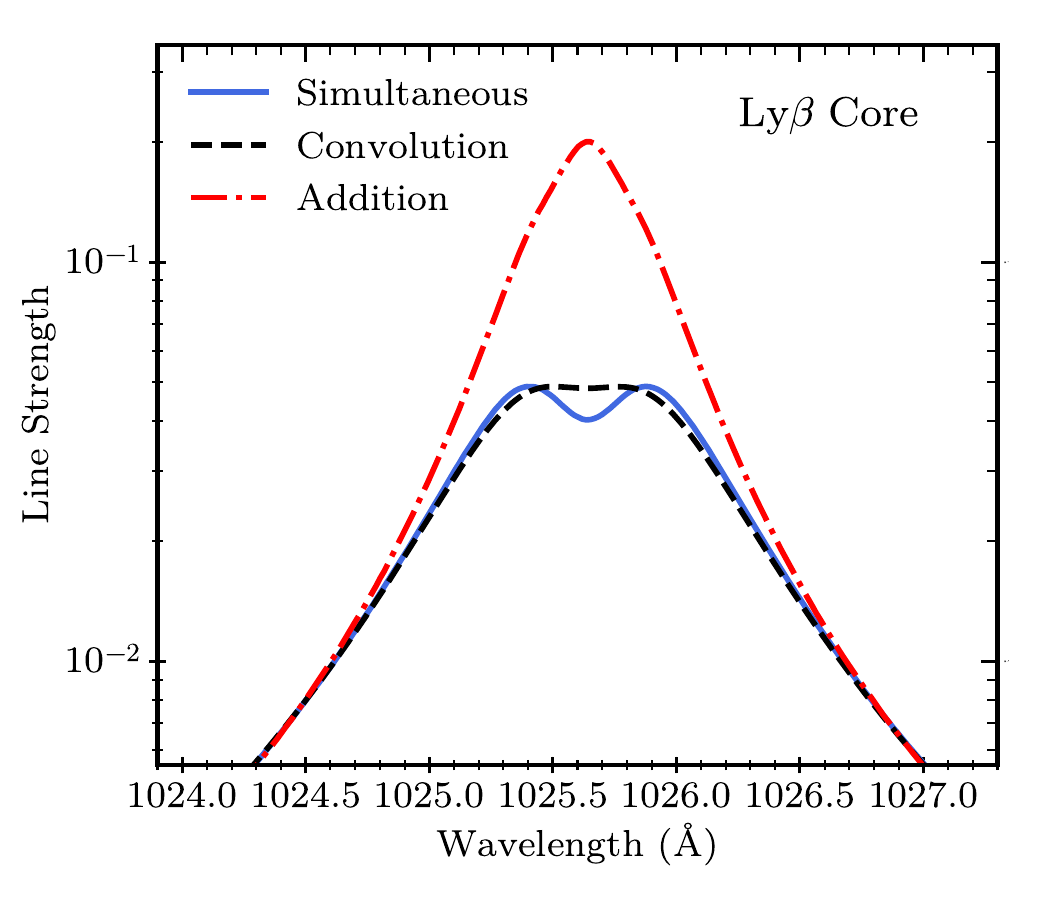}
    \caption{{\sc{Xenomorph}} quasi-H$_2^+$ Ly$\beta$ line shapes at $\textrm{T}=10,000~\textrm{K}$ and $n_e = 10^{17} $~cm$^{-3}$. The red dot-dash profile was calculated by adding separate ion-only and electron-only profiles together. The black dashed line was calculated by convolving those same ion-only and electron-only profiles together. The solid blue profile was calculated with simultaneous electron and ion broadening, by including both types of perturbers in the same simulation.}
    \label{fig:convolution_v_add}
\end{figure}

In Fig.~\ref{fig:convolution_v_add} we compare these different methods of incorporating ion and electron broadening for a Ly$\beta$ profile. In the static limit, out in the far wing, all three methods (a simple addition, a convolution, and simultaneous broadening) are equivalent (when ignoring the normalization of the simple addition), but the cores are drastically different. The simple addition naturally doubles the total opacity and fails to conserve oscillator strength. The true convolution preserves the oscillator strength, but loses the expected double peaked structure in the line core.

\subsection{Choice of Critical Radius}
\label{sec:crit_radius_choice}

\begin{figure}
    \centering
    \includegraphics[width=\columnwidth]{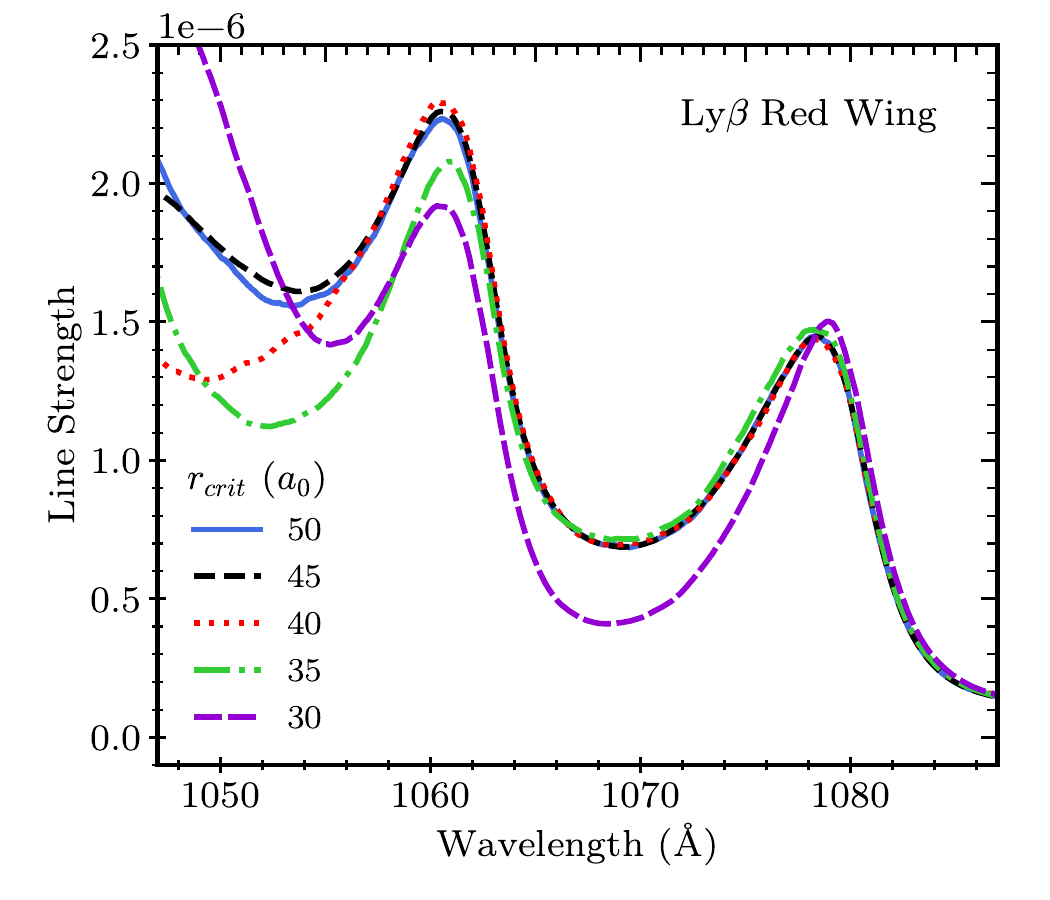}
    \caption{Convergence of the 1060Å and 1080Å Ly$\beta$ quasi-H$^+_2$ features with varying critical radii.}
    \label{fig:convergence}
\end{figure}
A critical radius must be chosen to determine when to operate in an atomic basis and when to operate in a molecular basis. The ideal critical radius can in general depend on many factors, including the quasi-molecular species, the transitions considered, the multipole order used in calculating $V(t)$, the directional correlation prescription, and the screening prescription. 
We caution that the exact choice is, however, still somewhat arbitrary.
We do not attempt to provide a general answer for what the ideal choice of critical radius is in this work.
This will be a focus of future efforts.
The primary quasi-H$_2^+$ features we consider here are the 1060Å and 1080Å Ly$\beta$ red wing resonances, as well as the 1400Å Ly$\alpha$ red wing resonance. 
The 1060Å resonance is formed at the largest internuclear separations of the three, as can be seen in Figs.~\ref{fig:h2p_curve} and \ref{fig:convergence}, and therefore requires the largest critical radius. The line shapes presented in this work use $r_{\mathrm{crit}} = 45 \;a_0$ or $r_{\mathrm{crit}} = 50 \;a_0$ unless otherwise stated.

\begin{figure}
    \centering
    \includegraphics[width=\linewidth]{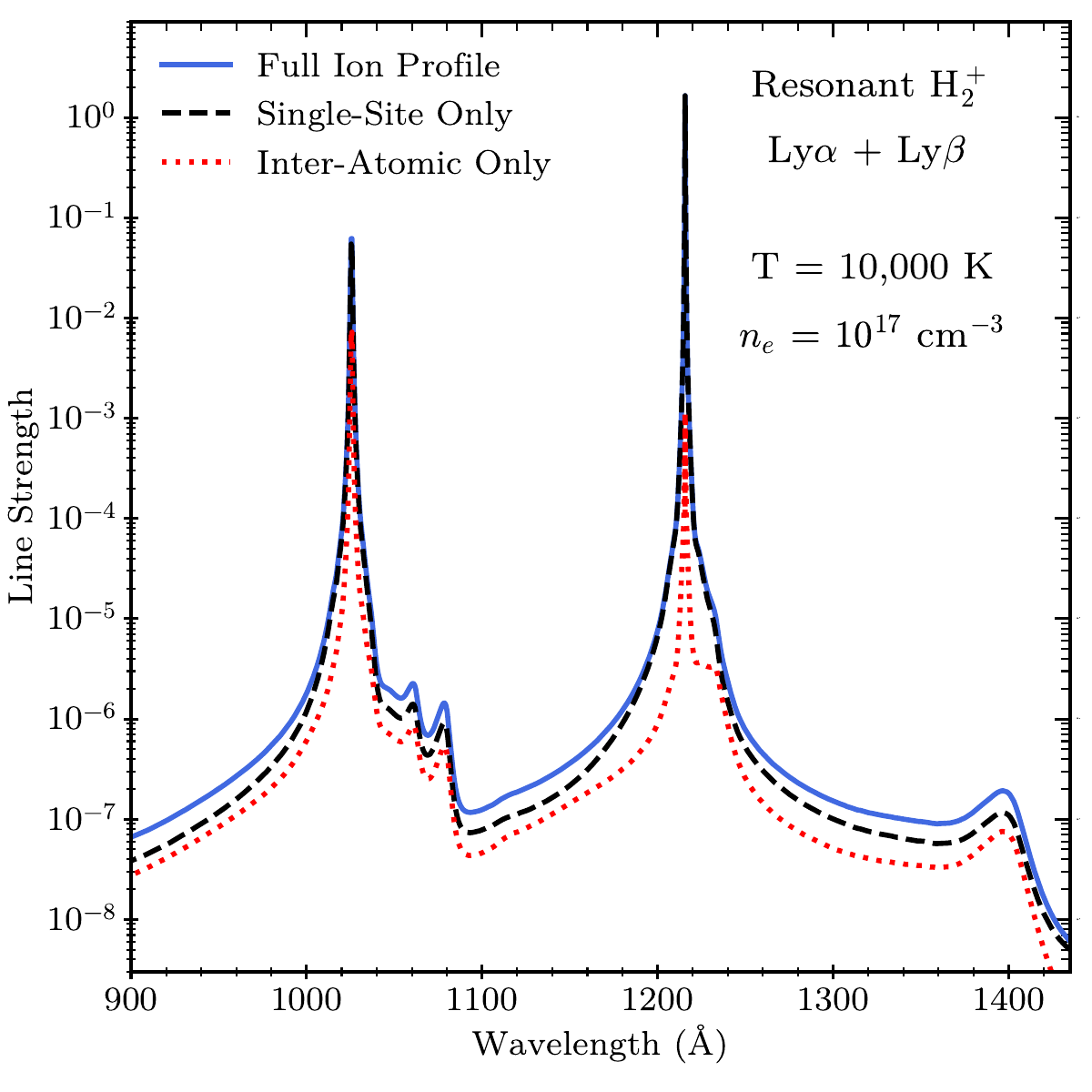}
    \caption{A Ly$\alpha$ and Ly$\beta$ line shape split between the inter-atomic contribution and the single-site contribution, as defined in Sec.~\ref{sec:inter_atomic_results}}
    \label{fig:interatomic_transitions}
\end{figure}

\subsection{Inter-Atomic Transitions}
\label{sec:inter_atomic_results}

In Sec.~\ref{sec:line_definition} and App.~\ref{app:unitarity} we discuss how inter-atomic transitions naturally arise from our multi-basis line shape method. 
Here we show why they must be included in our final line shape calculations.
The power spectrum method line shape equation (Eq.~\ref{eq:fls_power}) depends on the Fourier transform of the time evolved dipole operator $\vec D_{fi}(\omega)$, where $i$ and $f$ denote initial and final states respectively.
$\vec D(\omega)$ can be written as a block matrix equation with both single-site and inter-atomic transitions included following Eq.~\eqref{eq:Dt_block_form},
\begin{equation}
    D^a(\omega) = \begin{bmatrix}
D_{rr}(\omega) & D_{rp}(\omega) \\
D_{pr}(\omega) & D_{pp}(\omega)
\end{bmatrix},
\label{eq:Dt_block_form_iat}
\end{equation}
where the single-site block matrix $D_{rr}(\omega)$ and the inter-atomic block matrix $D_{pr}(\omega)$ both contribute to our final line shape profile. \red{Note that $D_{pp}(\omega)$ and $D_{rp}(\omega)$ do not contribute to the final profile, following App.~\ref{app:preserving_mu}}. 
In Fig.~\ref{fig:interatomic_transitions} we show a combined \textsc{Xenomorph} Ly$\alpha$ and Ly$\beta$ profile and separate out these inter-atomic and single-site contributions. 
The single-site profile (from $D_{rr}(\omega)$) dominates the line core while the inter-atomic profile (from $D_{pr}(\omega)$) is more comparable in strength in the line wings, and serves to raise the far wing relative to the line core. 
This is expected, as the inter-atomic transitions arise because of charge exchange, which occurs during close collisions. 
To be explicit, this difference between the profiles means we cannot ignore the inter-atomic contribution without distorting the strength of the line wing relative to the line core.

\section{Stellar Atmospheres Impact}
\label{sec:stellar_atmospheres_impact}

\begin{figure}
    \centering
    \includegraphics[width=\linewidth]{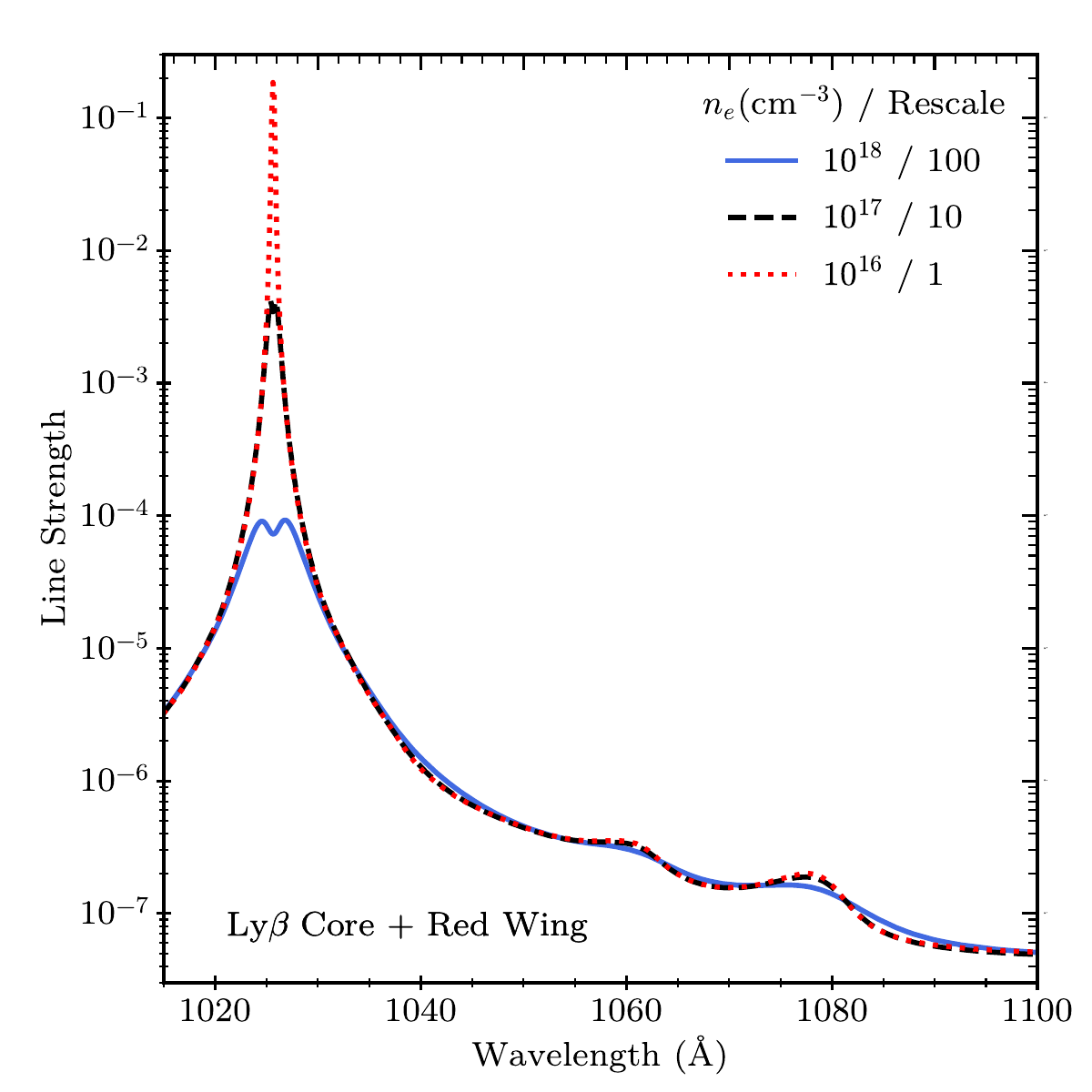}
    \caption{Convergence of Ly$\beta$ far line wing with varying density. Each line profile is area normalized and rescaled based on the relative electron density. The temperature (20,000 K) is fixed for each profile.}
    \label{fig:density_convergence}
\end{figure}

\begin{figure}
    \centering
    \includegraphics[width=\linewidth]{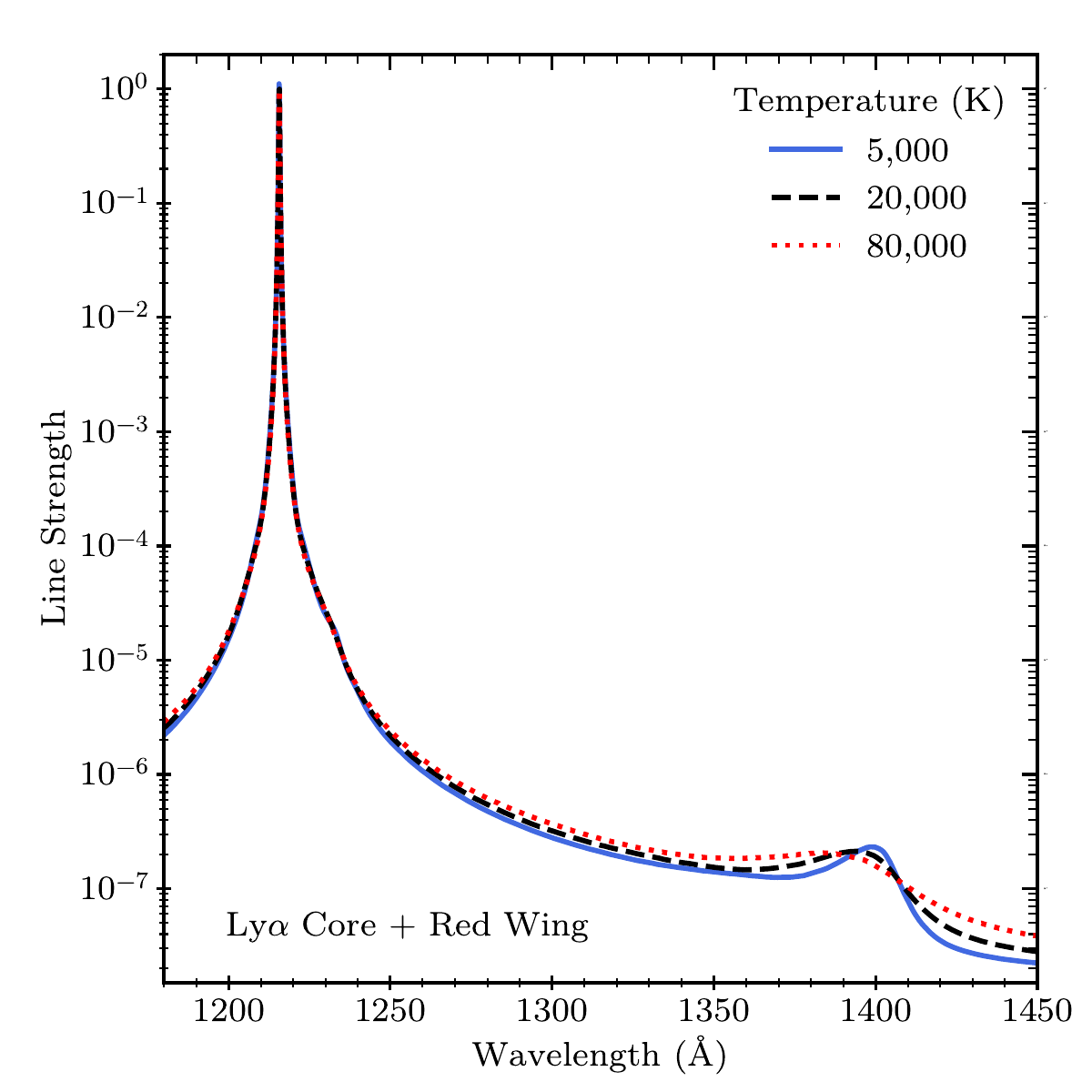}
    \caption{Differences in Ly$\alpha$ far line wing with varying temperature. The density ($n_e = 10^{17}$ cm$^{-3}$) is fixed for each profile. }
    \label{fig:temp_dependence}
\end{figure}

\red{We have implemented a grid of our new \textsc{Xenomorph} Ly$\alpha$ and Ly$\beta$ profiles, using simultaneous ion and electron broadening \citep{Cho2022}, into the stellar atmosphere and synthetic spectra code \textsc{Tlusty} \citep{Hubeny95,Hubeny21} version 211i. This is the same version of \textsc{Tlusty} that was previously used in \citet{Gomez2025}, which was modified to incorporate the new line-broadening data.}
The grid includes six logarithmically spaced densities,  $\textrm{log}(n_e) = 13, 14, ... 18$.
Line shape profiles at low densities ($n_e \leq 10^{16}\, \textrm{cm}^{-3}$) are approximated by rescaling an $n_e = 10^{17} \, \textrm{cm}^{-3}$ profile. This is a reasonable approximation because the shape of the far wing converges at low densities, as demonstrated in Fig.~\ref{fig:density_convergence}. 
We utilize this approximation because low-density simulations are more computationally expensive;  
an order-of-magnitude decrease in density makes close collisions an order-of-magnitude less common and results in the simulations having to run for an order-of-magnitude longer in physical time. 
Developing techniques to accelerate the convergence of the far line wing in low-density line shape simulations is a goal of future work.
We include three temperatures in this grid, T($10^3\,$K)$= 5, 20, 80 $, as demonstrated in Fig.~\ref{fig:temp_dependence}.

\begin{figure}
    \centering
    \includegraphics[width=\linewidth]{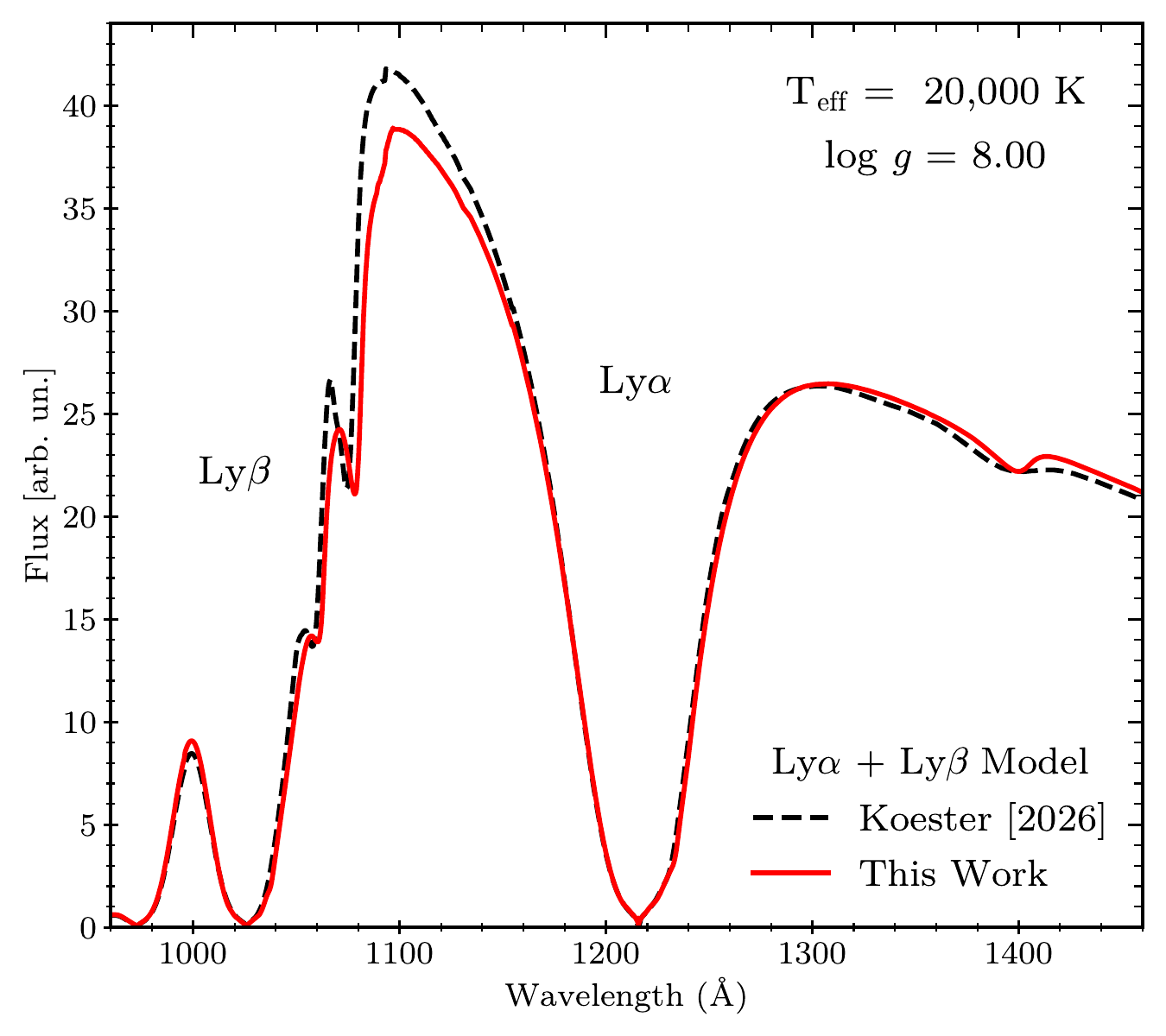}
    \caption{Model hydrogen-atmosphere white dwarf star spectra calculated with the \textsc{Tlusty} code with two different prescriptions for the Ly$\alpha$ and Ly$\beta$ line shape profiles, including both ion and electron broadening. The temperature, surface gravity, and all other atmosphere input parameters are held constant besides the Ly$\alpha$ and Ly$\beta$ profiles. The two model spectra are normalized such that total integrated flux over all frequencies is equal.}
    \label{fig:tlusty_koester_comp}
\end{figure}

\begin{figure}
    \centering
    \includegraphics[width=\linewidth]{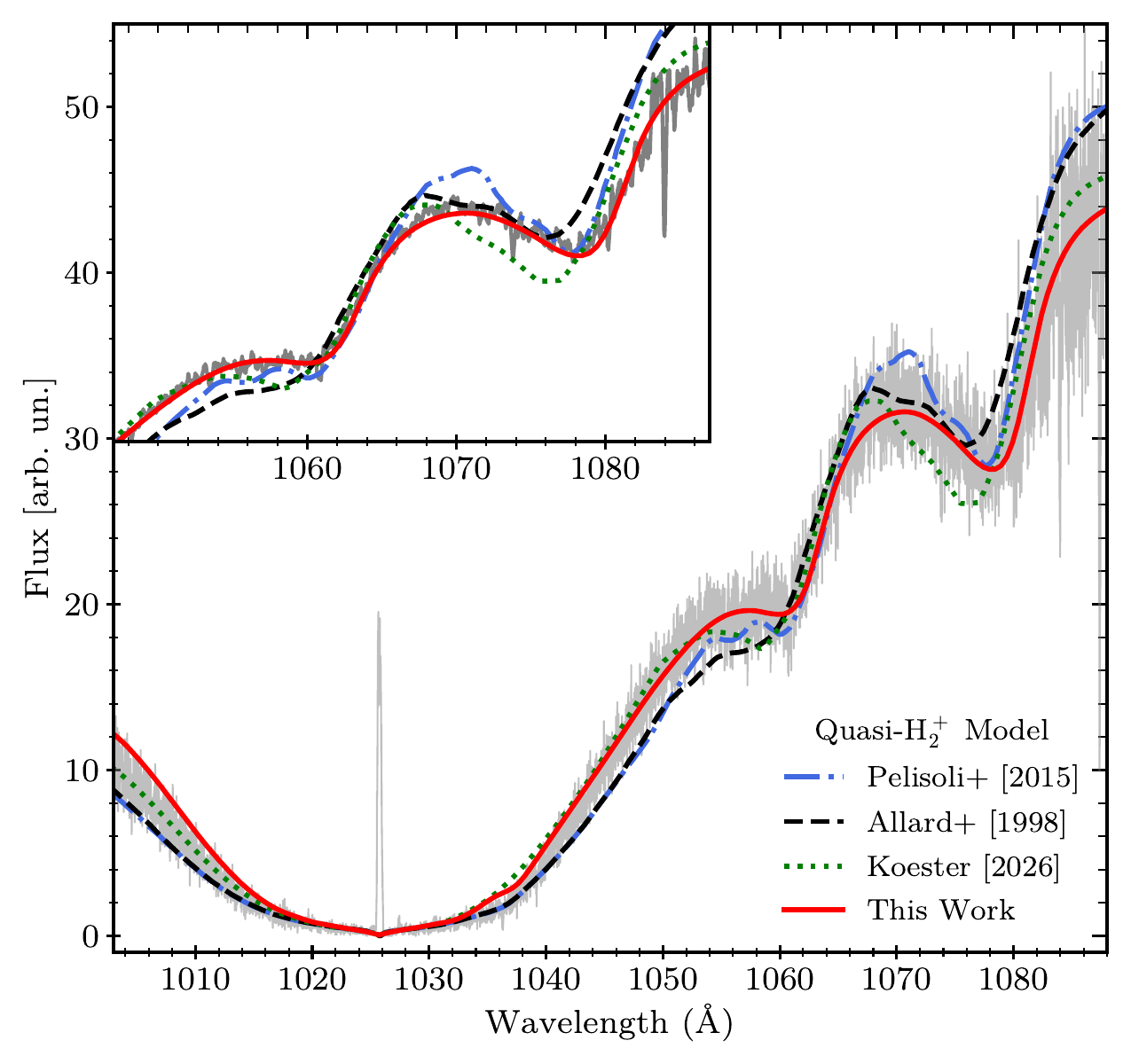}
    \caption{Model spectra comparison to an observed FUSE spectrum of hydrogen-atmosphere white dwarf star Wolf 1346, near the hydrogen Ly$\beta$ line. The data is shown in gray. The solid red model uses our new \textsc{Xenomorph} line shapes, while the dotted green model uses the line shapes from \citet{KoesterPrivate}. We generate both of these new model spectra with \textsc{Tlusty} version 211i. We also compare to two model spectra from Fig.~7 of \citet{Pelisoli15} that were generated with the \textsc{Tlusty} and \textsc{Synspec} \citep{Hubeny2011} codes: The dot-dash blue curve uses line shapes from \citet{Pelisoli15}, while the dashed black curve uses line shapes from \citet{Allard1998_1}. 
    We normalize all four models to minimize the model-data $\chi^2$ between 1040 and 1085 Å.
    The top left inset focuses on the 1060 and 1080 Å features, and smooths both the models and the data with a 0.2 Å boxcar average.
    \red{The FUSE spectrum for Wolf 1346 is available in MAST:\dataset[10.17909/exjv-h446]{https://doi.org/10.17909/exjv-h446}}.
    }
    \label{fig:tlusty_wolf_comp}
\end{figure}

In Fig.~\ref{fig:tlusty_koester_comp} we compare two DA WD synthetic spectra, each calculated with \textsc{Tlusty}, but with different quasi-H$_2^+$ Ly$\alpha$ and Ly$\beta$ profiles. The dashed black curve uses updated ULBT Ly$\alpha$ and Ly$\beta$ lines \citep{KoesterPrivate} with molecular data from \citet{Santos12,Pelisoli15}.
The solid red curve uses our new grid of Ly$\alpha$ and Ly$\beta$ profiles. 
The most significant model differences appear between Ly$\alpha$ and Ly$\beta$, and in the shape of the quasi-molecular features, which is consistent with our previous line shape comparisons in Figs.~\ref{fig:pelisoli_comp_wide} and~\ref{fig:pelisoli_panel_comp}.
The Ly$\beta$ red wing is shown in Fig.~\ref{fig:tlusty_wolf_comp} to highlight these differences in greater detail and demonstrate how they compare to an observed spectrum.

\red{In Fig.~\ref{fig:tlusty_wolf_comp} we compare synthetic DA spectra to a FUSE spectrum \citep{Moos2000,Sahnow2000} of WD star Wolf 1346 \citep{Holberg01,Hebrard2003}.}
As in \citet{Pelisoli15}, this is only intended to be a qualitative comparison.
All the model spectra in Fig.~\ref{fig:tlusty_wolf_comp} were calculated at the nominal conditions of T$_{\textrm{eff}} = 20,700$ K and log $g$ = 8.00, as estimated by \citet{Giammichele2012} using fits to optical hydrogen Balmer lines. 
The solid red and dotted green curves are new \textsc{Tlusty} model spectra that use line shape profiles from this work (solid red) and from \citet{KoesterPrivate} (dotted green). 
We also compare to \textsc{Tlusty} and \textsc{Synspec} model spectra from \citet{Pelisoli15} that were calculated with line shape profiles from \citet{Pelisoli15} (dot-dash blue) and \citet{Allard1998_1} (dashed black).
Every model plotted in Fig.~\ref{fig:tlusty_wolf_comp} has been normalized to minimize the model-data $\chi^2$ in the Ly$\beta$ red wing. 

We find good qualitative agreement in Fig.~\ref{fig:tlusty_wolf_comp} between the data and the model spectrum with our new multi-basis quasi-molecular line shape profiles.
The reduced model flux between Ly$\alpha$ and Ly$\beta$, evident in Fig.~\ref{fig:tlusty_koester_comp}, is consistent with the observed spectrum and brings the slope of the Ly$\beta$ red wing into better agreement.
Our smoother quasi-molecular profiles agree well with the observed shape of the two Ly$\beta$ quasi-H$_2^+$ resonances, without needing any additional broadening.
\red{These differences in model spectra are also consistent with the line shape profile differences we found in Fig.~\ref{fig:pelisoli_panel_comp}.}
Finally, the redshift in the quasi-H$_2^+$ resonances compared to the \citet{KoesterPrivate} and \citet{Allard1998_1} profiles also appears to match the data.
A more robust quantitative comparison to many ultraviolet white dwarf star spectra is outside the scope of this work and will be the focus of future efforts.

\section{Summary and Future Outlook}
\label{sec:conclusions}
We have presented a new approach to modeling line shapes with quasi-molecular resonances that are particularly relevant to hydrogen-atmosphere white dwarf stars, using a novel multi-basis procedure for simulation line shape calculations.
In this multi-basis method we repeatedly switch between an atomic basis and a molecular basis depending on the nearest-neighbor ion separation, while accounting for distant perturbers with atomic or molecular multipole moments. 
We have successfully implemented our multi-basis approach into the line shape code \textsc{Xenomorph} and demonstrated its ability to calculate hydrogen Lyman lines with quasi-H$_2^+$ features at hydrogen-atmosphere white dwarf star photosphere conditions.
We find that when applied to the model atmosphere and spectral synthesis code \textsc{Tlusty}, we can find good agreement with observed ultraviolet spectra of hydrogen-atmosphere white dwarf stars at conditions estimated from optical Balmer line fits.

This work represents the first simulation line shape code (Sec.~\ref{sec:Simulation_LS_Method}) implementation of quasi-molecular structure.
We caution, however, that this is still far from a perfect approach.
One current limitation is that this method still relies on a number of line shape theory approximations, many of which are discussed in Sec.~\ref{sec:results}, that could be improved upon in the future.
Another is that some physical effects, like screening and directional correlations are treated in an inconsistent manner depending on the basis used.
Finally, another remaining approximation is the $\mu$-ion model, as discussed in App.~\ref{app:preserving_mu}, which operates in the frame of reference of a single radiating atom.
These remaining issues are left for future work.

We have qualitatively demonstrated in Sec.~\ref{sec:stellar_atmospheres_impact} how our new line shapes can impact ultraviolet stellar spectra in hydrogen-atmosphere white dwarf stars.
However, comprehensive fits to many observed spectra, like in the work of \citet{Sahu2023}, are needed before robust claims can be made about how this work impacts unresolved discrepancies between ultraviolet and optical fits to hydrogen-atmosphere white dwarf stars, or to any of the other applications of ultraviolet white dwarf star model spectra discussed in Sec.~\ref{sec:intro}.
\red{Finally, while we have limited ourselves to quasi-H$_2^+$ Lyman lines in this work, we expect that higher-$n$ lines like the Balmer series, collisions involving heavier nuclei (Z$>$1), and multi-bound-electron systems (including resonance broadening) can be explored by extending the present multi-basis method.}
These subjects will also be the focus of future work.

\section*{Acknowledgments}
J.R.W., M.H.M., B.H.D., and D.E.W. acknowledge support from the Wootton Center for Astrophysical Plasma Properties, United States under U.S. Department of Energy Cooperative Agreement No. DE-NA0004149.
J.R.W. acknowledges support from the United States Department of Energy National Nuclear Security Administration SSGF program under DE-NA0003960.
LANL is operated by Triad National Security, LLC, for the National Nuclear Security Administration of the U.S. Department of Energy under Contract No. 89233218NCA000001. 
M.C.Z and J.R.W. would like to specifically acknowledge Los Alamos National Laboratory (LANL) ASC PEM Atomic Physics Project. 
D.V.F. and I.B. are supported by resources provided by the Australian Research Council, the Pawsey Supercomputing Centre, the National Computing Infrastructure of Australia, the Texas Advanced Computing Center, and the United States Air Force Office of Scientific Research.
We would also like to thank Prof. Detlev Koester for sharing his quasi-molecular data with us, and Prof. Pier-Emmanuel Tremblay for valuable discussions. 

\section*{Data Availability}

\red{The new quasi-molecular line shapes presented in this paper are available from the corresponding author (J.R.W.) upon request. Both the }\textsc{Tlusty} \red{and} \textsc{Synspec} \red{codes are currently undergoing a major upgrade to  modern Fortran 95. The data reported in this paper were implemented in an older,  Fortran 77, version of} \textsc{Tlusty}. \red{Both upgraded codes will be released soon. In the meantime, the most recent versions of the codes are available from I.H. upon request.}

\clearpage
\newpage

\appendix

\section{Inter Atomic Matrix Elements}
\label{app:unitarity}
Following a close collision, we perform a change of basis $U^m \rightarrow U^a$ in Eq.~\eqref{eq:secondCOB}.
As discussed in Sec.~\ref{sec:unitarity}, our atomic basis includes both radiator and perturber atomic states. 
$U^a(t)$ can therefore be written in block matrix form,
\begin{equation}
        U(t) = \begin{bmatrix}
            U_{rr}(t) & U_{rp}(t) \\
            U_{pr}(t) & U_{pp}(t)
        \end{bmatrix},
        \label{eq:u_block}
\end{equation}
where $U_{rp}(t)$ denotes the square sub-matrix with a perturber initial state and a radiator final state. Note that we have dropped the explicit $^a$ from $U^a(t)$, but every equation in the appendices should still be understood as having a basis of atomic states.
$\vec D$ can likewise be written as a block matrix,
\begin{equation}
    \vec D = \begin{bmatrix}
    \vec D_r & 0 \\
    0 & \vec D_p \\
\end{bmatrix},
\end{equation} 
where $\vec D_r$ and $\vec D_p$ represent the unperturbed atomic dipole matrix for the radiator and perturber, respectively. 
$H(t)$ can be written in a similar fashion,
\begin{equation}
    H(t) = \begin{bmatrix}
        H_r(t) & 0 \\
        0 & H_p(t) \\
    \end{bmatrix}.
    \label{eq:block_diagonal_hamil}
\end{equation}
The time evolved dipole matrix $\vec D(t)$ can then be written out in terms of individual block matrix partitions,
\begin{equation}
    \vec D(t) = U^\dag(t) \vec D U(t) = \begin{bmatrix}
        U_{rr}^\dag \vec D_r U_{rr} + U_{pr}^\dag \vec D_p U_{pr} & U^\dag_{rr} \vec D_r U_{rp} + U_{pr}^\dag \vec D_p U_{pp} \\
        U^\dag_{rp} \vec D_r U_{rr} + U^\dag_{pp} \vec D_p U_{pr} & U_{rp}^\dag \vec D_r U_{rp} + U_{pp}^\dag \vec D_p U_{pp}
    \end{bmatrix},
    \label{eq:long_dt}
\end{equation}
where we have suppressed the explicit time dependence of $U(t)$ for brevity.
If the off-diagonal blocks of $U(t)$ are zero, $\vec D(t)$ reduces to the simple case of two separate atomic systems,
\begin{equation}
    U(t) = \begin{bmatrix}
        U_{rr}(t) & 0 \\
        0 & U_{pp}(t)
    \end{bmatrix} \rightarrow
    \vec D(t) = \begin{bmatrix}
        U_{rr}^\dag \vec D_r U_{rr} & 0 \\
        0 & U_{pp}^\dag \vec D_p U_{pp}
    \end{bmatrix},
\end{equation}
either of which could be modeled in the standard $\mu$-ion approach.
However, the off-diagonal blocks of $U(t)$ are non-zero following a close collision, due to charge exchange, so we must instead use Eq.~\eqref{eq:long_dt}. 
As a reminder, $U(t)$ is updated (Eq.~\ref{time_evo_apx}) according to $H(t)$,
\begin{equation}
    U(t+\Delta t) \approx e^{-\textrm{i}H(t) \Delta t} U(t),
\end{equation}
which means that $U(t)$ now depends on both $H_r(t)$ and $H_p(t)$, following a close collision.
Expanding again in terms of individual block matrix partitions,
\begin{equation}
    U(t+\Delta t) \approx \begin{bmatrix}
        e^{-\textrm{i} H_r(t) \Delta t} U_{rr}(t) & e^{-\textrm{i} H_r(t) \Delta t} U_{rp}(t) \\
        e^{-\textrm{i} H_p(t) \Delta t} U_{pr}(t) & e^{-\textrm{i} H_p(t) \Delta t} U_{pp}(t)
    \end{bmatrix},
    \label{eq:time_evo_hrhp}
\end{equation}
we see that half the matrix depends on the Hamiltonian of the radiator, while the other half depends on the Hamiltonian of the perturber.

\section{Using the $\mu$-ion model}
\label{app:preserving_mu}
Calculating the Hamiltonian of the radiator $H_r(t)$ and the perturber $H_p(t)$, following a close collision, is challenging in the $\mu$-ion model for two reasons. First, the radiator is fixed at the center of the simulation sphere in the $\mu$-ion model, while the perturber is not. Second, additional close collisions involving the perturber \textit{or} the radiator will further expand the size of the atomic basis, and increase the computational cost of the simulation exponentially. 
In this appendix we describe how we avoid tracking the perturber Hamiltonian after a close collision ends, which allows us to continue using the $\mu$-ion model for the time being.

Utilizing the block matrix form of $\vec D(t)$ (Eq.~\ref{eq:long_dt}) and the linearity of a Fourier transform we can write $\vec D(\omega)$ as,
\begin{equation}
        \vec D(\omega) = \mathcal{F}\{\vec D(t)\} = \begin{bmatrix}
        \mathcal{F}\{U_{rr}^\dag \vec D_r U_{rr}\} + \mathcal{F}\{U_{pr}^\dag \vec D_p U_{pr}\} & \mathcal{F}\{U^\dag_{rr} \vec D_r U_{rp}\} + \mathcal{F}\{U_{pr}^\dag \vec D_p U_{pp}\} \\
        \mathcal{F}\{U^\dag_{rp} \vec D_r U_{rr}\} + \mathcal{F}\{U^\dag_{pp} \vec D_p U_{pr}\} & \mathcal{F}\{U_{rp}^\dag \vec D_r U_{rp}\} + \mathcal{F}\{U_{pp}^\dag \vec D_p U_{pp}\}
    \end{bmatrix},
    \label{eq:long_dw}
\end{equation}
where the line shape equation (Eq.~\ref{eq:fls_power}) can be written as,
\begin{equation}
    I(\omega) = \sum_{if} \rho_i |\vec D_{fi}(\omega)|^2,
\end{equation}
with subscripts $i$ and $f$ denoting initial and final states, respectively. \red{To be clear, the sums over $i$ and $f$ each include both `perturber' and `radiator' states.} 
In the $\mu$-ion model we assume the density matrix $\rho$ is diagonal and non-zero only for radiator states.
We therefore only need $\vec D_r(\omega)$, the sub-matrix where the initial state is a radiator state,
\begin{equation}
        \vec D_r(\omega) = \begin{bmatrix}
        \mathcal{F}\{U_{rr}^\dag \vec D_r U_{rr}\} + \mathcal{F}\{U_{pr}^\dag \vec D_p U_{pr}\} \\ \mathcal{F}\{U^\dag_{rp} \vec D_r U_{rr}\} + \mathcal{F}\{U_{pp}^\dag \vec D_p U_{pr}\}
    \end{bmatrix}.
    \label{eq:dw_r}
\end{equation}
$\vec D_r(\omega)$ thus depends on both $H_r(t)$ (through $U_{rr}$ and $U_{rp}$) and $H_p(t)$ (through $U_{pp}$ and $U_{pr}$). Here we make the following approximations,
\begin{equation}
    \mathcal{F}\{U_{pr}^\dag \vec D_p U_{pr}\} \approx \mathcal{F}\{U_{rp}^\dag \vec D_r U_{rp}\},
    \label{dw_approx1}
\end{equation}
\begin{equation}
    \mathcal{F}\{U_{pp}^\dag \vec D_p U_{pr}\} \approx \mathcal{F}\{U_{rr}^\dag \vec D_r U_{rp}\}.
    \label{dw_approx2}
\end{equation}
We justify these approximations on the basis of the symmetry of the quasi-molecule at hand. Both the radiator and the perturber have the same nuclear charge (Z = 1) so their one-electron atomic dipoles are equivalent, $\vec D_r$ = $\vec D_p$.  
They are also both moving through the same plasma so their \textit{time-averaged} power spectrum should be equivalent. 
Note that the density matrix never enters into our calculation of $\vec D(\omega)$, so our assumptions about $\rho$ do not disrupt the symmetry that justifies Eqs.~\eqref{dw_approx1} and \eqref{dw_approx2}.
These approximations then allow us to rewrite $\vec D_r(\omega)$,
\begin{equation}
    \vec D_r(\omega) \approx \begin{bmatrix}
        \mathcal{F}\{U_{rr}^\dag \vec D_r U_{rr}\} + \mathcal{F}\{U_{rp}^\dag \vec D_r U_{rp}\} \\ \mathcal{F}\{U_{rp}^\dag \vec D_r U_{rr}\} + \mathcal{F}\{U^\dag_{rr} \vec D_r U_{rp}\}
    \end{bmatrix},
\end{equation}
such that $\vec D_r(\omega)$ now exclusively depends on $H_r$ and not $H_p$. This allows us to continue using our $\mu$-ion model, where $H_r(t)$ is the only atomic Hamiltonian we track.
An improved model beyond the $\mu$-ion approximation, which tracks the time history of many different atoms for the entire simulation, would eliminate the need to make the approximations in Eqs.~\eqref{dw_approx1} and \eqref{dw_approx2}. Such a model is a goal of future work.

\newpage
\bibliography{h2pqm_rev}{}

@ARTICLE{Zammit2017,
       author = {{Zammit}, Mark C. and {Savage}, Jeremy S. and {Colgan}, James and {Fursa}, Dmitry V. and {Kilcrease}, David P. and {Bray}, Igor and {Fontes}, Christopher J. and {Hakel}, Peter and {Timmermans}, Eddy},
        title = "{State-resolved Photodissociation and Radiative Association Data for the Molecular Hydrogen Ion}",
      journal = {\apj},
         year = 2017,
        month = dec,
       volume = {851},
       number = {1},
          eid = {64},
        pages = {64},
          doi = {10.3847/1538-4357/aa9712},
       adsurl = {https://ui.adsabs.harvard.edu/abs/2017ApJ...851...64Z}
}

@ARTICLE{Gomez2021,
       author = {{Gomez}, T.~A. and {Nagayama}, T. and {Cho}, P.~B. and {Zammit}, M.~C. and {Fontes}, C.~J. and {Kilcrease}, D.~P. and {Bray}, I. and {Hubeny}, I. and {Dunlap}, B.~H. and {Montgomery}, M.~H. and {Winget}, D.~E.},
        title = "{All-Order Full-Coulomb Quantum Spectral Line-Shape Calculations}",
      journal = {\prl},
         year = 2021,
        month = dec,
       volume = {127},
       number = {23},
          eid = {235001},
        pages = {235001},
          doi = {10.1103/PhysRevLett.127.235001},
       adsurl = {https://ui.adsabs.harvard.edu/abs/2021PhRvL.127w5001G}
}

@ARTICLE{Gomez2022,
       author = {{Gomez}, T.~A. and {Nagayama}, T. and {Cho}, P.~B. and {Kilcrease}, D.~P. and {Fontes}, C.~J. and {Zammit}, M.~C.},
        title = "{Introduction to spectral line shape theory}",
      journal = {Journal of Physics B Atomic Molecular Physics},
         year = 2022,
        month = feb,
       volume = {55},
       number = {3},
          eid = {034002},
        pages = {034002},
          doi = {10.1088/1361-6455/ac4f31},
       adsurl = {https://ui.adsabs.harvard.edu/abs/2022JPhB...55c4002G}
}

@ARTICLE{Cho2022,
       author = {{Cho}, P.~B. and {Gomez}, T.~A. and {Montgomery}, M.~H. and {Dunlap}, B.~H. and {Fitz Axen}, M. and {Hobbs}, B. and {Hubeny}, I. and {Winget}, D.~E.},
        title = "{Simulation of Stark-broadened Hydrogen Balmer-line Shapes for DA White Dwarf Synthetic Spectra}",
      journal = {\apj},
         year = 2022,
        month = mar,
       volume = {927},
       number = {1},
          eid = {70},
        pages = {70},
          doi = {10.3847/1538-4357/ac4df3},
       adsurl = {https://ui.adsabs.harvard.edu/abs/2022ApJ...927...70C}
}

@ARTICLE{Allard1999,
       author = {{Allard}, N.~F. and {Royer}, A. and {Kielkopf}, J.~F. and {Feautrier}, N.},
        title = "{Effect of the variation of electric-dipole moments on the shape of pressure-broadened atomic spectral lines}",
      journal = {\pra},
         year = 1999,
        month = aug,
       volume = {60},
       number = {2},
        pages = {1021-1033},
          doi = {10.1103/PhysRevA.60.1021},
       adsurl = {https://ui.adsabs.harvard.edu/abs/1999PhRvA..60.1021A}
}

@ARTICLE{Schaeuble2019,
       author = {{Schaeuble}, M. -A. and {Nagayama}, T. and {Bailey}, J.~E. and {Gomez}, T.~A. and {Montgomery}, M.~H. and {Winget}, D.~E.},
        title = "{H{\ensuremath{\beta}} and H{\ensuremath{\gamma}} Absorption-line Profile Inconsistencies in Laboratory Experiments Performed at White Dwarf Photosphere Conditions}",
      journal = {\apj},
         year = 2019,
        month = nov,
       volume = {885},
       number = {1},
          eid = {86},
        pages = {86},
          doi = {10.3847/1538-4357/ab479d},
       adsurl = {https://ui.adsabs.harvard.edu/abs/2019ApJ...885...86S}
}

@ARTICLE{Gigosos87,
       author = {{Gigosos}, M.~A. and {Cardenoso}, V.},
        title = "{Study of the effects of ion dynamics on Stark profiles of Balmer-{\ensuremath{\alpha}} and -{\ensuremath{\beta}} lines using simulation techniques}",
      journal = {Journal of Physics B Atomic Molecular Physics},
         year = 1987,
        month = nov,
       volume = {20},
       number = {22},
        pages = {6005-6019},
          doi = {10.1088/0022-3700/20/22/013},
       adsurl = {https://ui-adsabs-harvard-edu.ezproxy.lib.utexas.edu/abs/1987JPhB...20.6005G}
}

@ARTICLE{Stambulchik06,
       author = {{Stambulchik}, E. and {Maron}, Y.},
        title = "{A study of ion-dynamics and correlation effects for spectral line broadening in plasma: K-shell lines}",
      journal = {\jqsrt},
         year = 2006,
        month = may,
       volume = {99},
       number = {1-3},
        pages = {730-749},
          doi = {10.1016/j.jqsrt.2005.05.058},
       adsurl = {https://ui.adsabs.harvard.edu/abs/2006JQSRT..99..730S}
}

@ARTICLE{Tremblay2020,
       author = {{Tremblay}, Patrick and {Beauchamp}, A. and {Bergeron}, P.},
        title = "{New Calculations of Stark-broadened Profiles for Neutral Helium Lines Using Computer Simulations}",
      journal = {\apj},
         year = 2020,
        month = oct,
       volume = {901},
       number = {2},
          eid = {104},
        pages = {104},
          doi = {10.3847/1538-4357/abb0e5},
archivePrefix = {arXiv},
       eprint = {2008.09834},
 primaryClass = {astro-ph.SR},
       adsurl = {https://ui.adsabs.harvard.edu/abs/2020ApJ...901..104T}
}

@ARTICLE{Hummer1988,
       author = {{Hummer}, D.~G. and {Mihalas}, Dimitri},
        title = "{The Equation of State for Stellar Envelopes. I. an Occupation Probability Formalism for the Truncation of Internal Partition Functions}",
      journal = {\apj},
         year = 1988,
        month = aug,
       volume = {331},
        pages = {794},
          doi = {10.1086/166600},
       adsurl = {https://ui.adsabs.harvard.edu/abs/1988ApJ...331..794H}
}

@INPROCEEDINGS{Zammit2018,
       author = {{Zammit}, M.~C. and {Savage}, J.~S. and {Colgan}, J. and {Fursa}, D.~V. and {Bray}, I. and {Leiding}, J. and {Nisoli}, C. and {Fontes}, C.~J. and {Kilcrease}, D.~P. and {Hakel}, P. and {Timmermans}, E.},
        title = "{The Los Alamos National Laboratory Molecular Opacity Project and the Photodissociation Isotopic Effects of H$_{2}$$^{+}$ and D$_{2}$$^{+}$}",
    booktitle = {Workshop on Astrophysical Opacities},
         year = 2018,
       series = {Astronomical Society of the Pacific Conference Series},
       volume = {515},
        month = aug,
        pages = {145},
       adsurl = {https://ui.adsabs.harvard.edu/abs/2018ASPC..515..145Z}
}

@ARTICLE{Zammit2019,
       author = {{Zammit}, Mark C. and {Charlton}, Michael and {Jonsell}, Svante and {Colgan}, James and {Savage}, Jeremy S. and {Fursa}, Dmitry V. and {Kadyrov}, Alisher S. and {Bray}, Igor and {Forrey}, Robert C. and {Fontes}, Christopher J. and {Leiding}, Jeffery A. and {Kilcrease}, David P. and {Hakel}, Peter and {Timmermans}, Eddy},
        title = "{Laser-driven production of the antihydrogen molecular ion}",
      journal = {\pra},
         year = 2019,
        month = oct,
       volume = {100},
       number = {4},
          eid = {042709},
        pages = {042709},
          doi = {10.1103/PhysRevA.100.042709},
       adsurl = {https://ui.adsabs.harvard.edu/abs/2019PhRvA.100d2709Z}
}

@ARTICLE{Stambulchik2022,
       author = {{Stambulchik}, Evgeny and {Iglesias}, Carlos A.},
        title = "{Full radiator-perturber interaction in computer simulations of hydrogenic spectral line broadening by plasmas}",
      journal = {\pre},
         year = 2022,
        month = may,
       volume = {105},
       number = {5},
          eid = {055210},
        pages = {055210},
          doi = {10.1103/PhysRevE.105.055210},
       adsurl = {https://ui.adsabs.harvard.edu/abs/2022PhRvE.105e5210S}
}

@ARTICLE{Baschek1984,
       author = {{Baschek}, B. and {Koeppen}, J. and {Scholz}, M. and {Wehrse}, R. and {Heck}, A. and {Jaschek}, C. and {Jaschek}, M.},
        title = "{The ultraviolet (IUE) spectra of the lambda Bootis stars.}",
      journal = {\aap},
         year = 1984,
        month = feb,
       volume = {131},
        pages = {378-384},
       adsurl = {https://ui.adsabs.harvard.edu/abs/1984A&A...131..378B}
}

@ARTICLE{Nelan_Wegner1985,
       author = {{Nelan}, E.~P. and {Wegner}, G.},
        title = "{Identification of the 1400 and 1600 A features observed in the ultraviolet spectra of DA white dwarfs}",
      journal = {\apjl},
         year = 1985,
        month = feb,
       volume = {289},
        pages = {L31-L33},
          doi = {10.1086/184428},
       adsurl = {https://ui.adsabs.harvard.edu/abs/1985ApJ...289L..31N}
}

@ARTICLE{Koester1985,
       author = {{Koester}, D. and {Weidemann}, V. and {Zeidler-K.~T.}, E.~M. and {Vauclair}, G.},
        title = "{The explanation of the 1400 and 1600 A features in DA white dwarfs.}",
      journal = {\aap},
         year = 1985,
        month = jan,
       volume = {142},
        pages = {L5-L8},
       adsurl = {https://ui.adsabs.harvard.edu/abs/1985A&A...142L...5K}
}

@ARTICLE{Kielkopf1995,
       author = {{Kielkopf}, John F. and {Allard}, Nicole F.},
        title = "{Satellites on Lyman Alpha Due to H-H and H-H + Collisions}",
      journal = {\apjl},
         year = 1995,
        month = sep,
       volume = {450},
        pages = {L75},
          doi = {10.1086/309663},
       adsurl = {https://ui.adsabs.harvard.edu/abs/1995ApJ...450L..75K}
}

@ARTICLE{Allard_Koester_1992,
       author = {{Allard}, N.~F. and {Koester}, D.},
        title = "{Theoretical profiles of Lyman alpha satellites and application to synthetic spectra of DA white dwarfs.}",
      journal = {\aap},
         year = 1992,
        month = may,
       volume = {258},
        pages = {464-468},
       adsurl = {https://ui.adsabs.harvard.edu/abs/1992A&A...258..464A}
}

@ARTICLE{Allard2009,
       author = {{Allard}, N.~F. and {Noselidze}, I. and {Kruk}, J.~W.},
        title = "{New study of the quasi-molecular Lyman-{\ensuremath{\gamma}} satellites due to H-H$^{+}$ collisions}",
      journal = {\aap},
         year = 2009,
        month = nov,
       volume = {506},
       number = {2},
        pages = {993-997},
          doi = {10.1051/0004-6361/200912511},
       adsurl = {https://ui.adsabs.harvard.edu/abs/2009A&A...506..993A}
}

@ARTICLE{Allard1998_1,
       author = {{Allard}, N.~F. and {Drira}, I. and {Gerbaldi}, M. and {Kielkopf}, J. and {Spielfiedel}, A.},
        title = "{New study of the quasi-molecular Lyman alpha satellites due to H-H and H-H(+) collisions}",
      journal = {\aap},
         year = 1998,
        month = jul,
       volume = {335},
        pages = {1124-1129},
       adsurl = {https://ui.adsabs.harvard.edu/abs/1998A&A...335.1124A}
}

@ARTICLE{GenestBeaulieu2019,
       author = {{Genest-Beaulieu}, C. and {Bergeron}, P.},
        title = "{A Comprehensive Spectroscopic and Photometric Analysis of DA and DB White Dwarfs from SDSS and Gaia}",
      journal = {\apj},
         year = 2019,
        month = feb,
       volume = {871},
       number = {2},
          eid = {169},
        pages = {169},
          doi = {10.3847/1538-4357/aafac6},
       adsurl = {https://ui.adsabs.harvard.edu/abs/2019ApJ...871..169G}
}

@PHDTHESIS{Fuchs2017,
       author = {{Fuchs}, Joshua Thomas},
        title = "{Fundamental Properties of White Dwarfs Alone and in Binaries}",
       school = {University of North Carolina, Chapel Hill},
         year = 2017,
        month = jan,
       adsurl = {https://ui.adsabs.harvard.edu/abs/2017PhDT........20F}
}

@ARTICLE{Xu2019,
       author = {{Xu}, Siyi and {Dufour}, Patrick and {Klein}, Beth and {Melis}, Carl and {Monson}, Nathaniel N. and {Zuckerman}, B. and {Young}, Edward D. and {Jura}, Michael A.},
        title = "{Compositions of Planetary Debris around Dusty White Dwarfs}",
      journal = {\aj},
         year = 2019,
        month = dec,
       volume = {158},
       number = {6},
          eid = {242},
        pages = {242},
          doi = {10.3847/1538-3881/ab4cee},
archivePrefix = {arXiv},
       eprint = {1910.07197},
 primaryClass = {astro-ph.SR},
       adsurl = {https://ui.adsabs.harvard.edu/abs/2019AJ....158..242X}
}

@ARTICLE{Gansicke2012,
       author = {{G{\"a}nsicke}, B.~T. and {Koester}, D. and {Farihi}, J. and {Girven}, J. and {Parsons}, S.~G. and {Breedt}, E.},
        title = "{The chemical diversity of exo-terrestrial planetary debris around white dwarfs}",
      journal = {\mnras},
         year = 2012,
        month = jul,
       volume = {424},
       number = {1},
        pages = {333-347},
          doi = {10.1111/j.1365-2966.2012.21201.x},
archivePrefix = {arXiv},
       eprint = {1205.0167},
 primaryClass = {astro-ph.EP},
       adsurl = {https://ui.adsabs.harvard.edu/abs/2012MNRAS.424..333G}
}

@ARTICLE{Sahu2023,
       author = {{Sahu}, Snehalata and {G{\"a}nsicke}, Boris T. and {Tremblay}, Pier-Emmanuel and {Koester}, Detlev and {Hermes}, J.~J. and {Wilson}, David J. and {Toloza}, Odette and {Hoskin}, Matthew J. and {Farihi}, Jay and {Manser}, Christopher J. and {Redfield}, Seth},
        title = "{AHST COS ultra-violet spectroscopic survey of 311 DA white dwarfs - I. Fundamental parameters and comparative studies}",
      journal = {\mnras},
         year = 2023,
        month = sep,
          doi = {10.1093/mnras/stad2663},
archivePrefix = {arXiv},
       eprint = {2309.00239},
 primaryClass = {astro-ph.SR},
       adsurl = {https://ui.adsabs.harvard.edu/abs/2023MNRAS.tmp.2596S}
}

@ARTICLE{Allard1991,
       author = {{Allard}, N. and {Kielkopf}, J.},
        title = "{Temperature and density dependence of the Lyman-alpha line wing in hydrogen-rich dwarf atmospheres.}",
      journal = {\aap},
         year = 1991,
        month = feb,
       volume = {242},
        pages = {133},
       adsurl = {https://ui.adsabs.harvard.edu/abs/1991A&A...242..133A}
}

@ARTICLE{Allard2003,
       author = {{Allard}, N.~F. and {Allard}, F. and {Hauschildt}, P.~H. and {Kielkopf}, J.~F. and {Machin}, L.},
        title = "{A new model for brown dwarf spectra including accurate unified line shape theory for the Na I and K I resonance line profiles}",
      journal = {\aap},
         year = 2003,
        month = dec,
       volume = {411},
        pages = {L473-L476},
          doi = {10.1051/0004-6361:20031299},
       adsurl = {https://ui.adsabs.harvard.edu/abs/2003A&A...411L.473A}
}

@ARTICLE{Allard2014,
       author = {{Allard}, N.~F. and {Alekseev}, V.~A.},
        title = "{Collisional profiles of ionized calcium perturbed by helium}",
      journal = {Advances in Space Research},
         year = 2014,
        month = oct,
       volume = {54},
       number = {7},
        pages = {1248-1253},
          doi = {10.1016/j.asr.2013.09.027},
       adsurl = {https://ui.adsabs.harvard.edu/abs/2014AdSpR..54.1248A}
}

@ARTICLE{Royer1980,
       author = {{Royer}, Antoine},
        title = "{Shift, width, and asymmetry of pressure-broadened spectral lines at intermediate densities}",
      journal = {\pra},
         year = 1980,
        month = oct,
       volume = {22},
       number = {4},
        pages = {1625-1654},
          doi = {10.1103/PhysRevA.22.1625},
       adsurl = {https://ui.adsabs.harvard.edu/abs/1980PhRvA..22.1625R}
}

@ARTICLE{Allard1994,
       author = {{Allard}, N.~F. and {Koester}, D. and {Feautrier}, N. and {Spielfiedel}, A.},
        title = "{Free-free quasi-molecular absorption and satellites in Lyman-alpha due to collisions with H and H\^+\^.}",
      journal = {\aaps},
         year = 1994,
        month = dec,
       volume = {108},
        pages = {417-431},
       adsurl = {https://ui.adsabs.harvard.edu/abs/1994A&AS..108..417A}
}

@ARTICLE{Junkel2000,
       author = {{Junkel}, G.~C. and {Gunderson}, M.~A. and {Hooper}, C.~F., Jr. and {Haynes}, D.~A., Jr.},
        title = "{Full Coulomb calculation of Stark broadened spectra from multielectron ions: A focus on the dense plasma line shift}",
      journal = {\pre},
         year = 2000,
        month = oct,
       volume = {62},
       number = {4},
        pages = {5584-5593},
          doi = {10.1103/PhysRevE.62.5584},
       adsurl = {https://ui.adsabs.harvard.edu/abs/2000PhRvE..62.5584J}
}

@article{Gomez24_PenetratingIons,
  title = {Impact of penetrating collisions of plasma ions on spectral line shapes},
  author = {Gomez, Thomas A. and Stambulchik, Evgeny and White, Jackson},
  journal = {Phys. Rev. A},
  volume = {109},
  issue = {5},
  pages = {052804},
  numpages = {9},
  year = {2024},
  month = {May},
  publisher = {American Physical Society},
  doi = {10.1103/PhysRevA.109.052804},
  url = {https://link.aps.org/doi/10.1103/PhysRevA.109.052804}
}

@ARTICLE{Stambulchik07,
       author = {{Stambulchik}, E. and {Fisher}, D.~V. and {Maron}, Y. and {Griem}, H.~R. and {Alexiou}, S.},
        title = "{Correlation effects and their influence on line broadening in plasmas: Application to H $_{{\ensuremath{\alpha}}}$}",
      journal = {High Energy Density Physics},
         year = 2007,
        month = may,
       volume = {3},
       number = {1-2},
        pages = {272-277},
          doi = {10.1016/j.hedp.2007.02.021},
       adsurl = {https://ui.adsabs.harvard.edu/abs/2007HEDP....3..272S}
}

@ARTICLE{Bohlin20,
       author = {{Bohlin}, Ralph C. and {Hubeny}, Ivan and {Rauch}, Thomas},
        title = "{New Grids of Pure-hydrogen White Dwarf NLTE Model Atmospheres and the HST/STIS Flux Calibration}",
      journal = {\aj},
         year = 2020,
        month = jul,
       volume = {160},
       number = {1},
          eid = {21},
        pages = {21},
          doi = {10.3847/1538-3881/ab94b4},
archivePrefix = {arXiv},
       eprint = {2005.10945},
 primaryClass = {astro-ph.SR},
       adsurl = {https://ui.adsabs.harvard.edu/abs/2020AJ....160...21B}
}

@ARTICLE{Iglesias13,
       author = {{Iglesias}, Carlos A.},
        title = "{Efficient algorithms for Stark-Zeeman spectral line shape calculations}",
      journal = {High Energy Density Physics},
         year = 2013,
        month = dec,
       volume = {9},
       number = {4},
        pages = {737-744},
          doi = {10.1016/j.hedp.2013.08.003},
       adsurl = {https://ui.adsabs.harvard.edu/abs/2013HEDP....9..737I}
}

@ARTICLE{Hubeny95,
       author = {{Hubeny}, I. and {Lanz}, T.},
        title = "{Non-LTE Line-blanketed Model Atmospheres of Hot Stars. I. Hybrid Complete Linearization/Accelerated Lambda Iteration Method}",
      journal = {\apj},
         year = 1995,
        month = feb,
       volume = {439},
        pages = {875},
          doi = {10.1086/175226},
       adsurl = {https://ui.adsabs.harvard.edu/abs/1995ApJ...439..875H}
}

@article{Gomez18a,
	adsurl = {http://adsabs.harvard.edu/abs/2018Atoms...6...22G},
	author = {{Gomez}, T. and {Nagayama}, T. and {Fontes}, C. and {Kilcrease}, D. and {Hansen}, S. and {Montgomery}, M. and {Winget}, D.},
	doi = {10.3390/atoms6020022},
	journal = {Atoms},
	month = apr,
	pages = {22},
	title = {{Matrix Methods for Solving Hartree-Fock Equations in Atomic Structure Calculations and Line Broadening}},
	volume = 6,
	year = 2018
}

@ARTICLE{Pelisoli15,
       author = {{Pelisoli}, Ingrid and {Santos}, M.~G. and {Kepler}, S.~O.},
        title = "{Unified line profiles for hydrogen perturbed by collisions with protons: satellites and asymmetries}",
      journal = {\mnras},
         year = 2015,
        month = apr,
       volume = {448},
       number = {3},
        pages = {2332-2343},
          doi = {10.1093/mnras/stv167},
archivePrefix = {arXiv},
       eprint = {1501.05609},
 primaryClass = {astro-ph.SR},
       adsurl = {https://ui.adsabs.harvard.edu/abs/2015MNRAS.448.2332P}
}

@ARTICLE{Calisti14,
       author = {{Calisti}, Annette and {Demura}, Alexander and {Gigosos}, Marco and {Gonz{\'a}lez-Herrero}, Diego and {Iglesias}, Carlos and {Lisitsa}, Valery and {Stambulchik}, Evgeny},
        title = "{Influence of Microfield Directionality on Line Shapes}",
      journal = {Atoms},
         year = 2014,
        month = jun,
       volume = {2},
       number = {2},
        pages = {259-276},
          doi = {10.3390/atoms2020259},
       adsurl = {https://ui.adsabs.harvard.edu/abs/2014Atoms...2..259C}
}

@article{Demura77, title={Effect of reduced mass in Stark broadening of hydrogen lines}, volume={46}, ISSN={1063-7761}, note={ADS Bibcode: 1977JETP...46..209D}, journal={Soviet Journal of Experimental and Theoretical Physics}, author={Demura, A. V. and Lisitsa, V. S. and Sholin, G. V.}, year={1977}, month=aug, pages={209} }

@article{Stambulchik16, title={Dynamic Stark broadening of Lyman- α}, volume={49}, ISSN={0953-4075, 1361-6455}, DOI={10.1088/0953-4075/49/3/035701}, number={3}, journal={Journal of Physics B: Atomic, Molecular and Optical Physics}, author={Stambulchik, Evgeny and Demura, Alexander V}, year={2016}, month=feb, pages={035701} }

@article{Kilcrease93, title={Ion broadening of dense-plasma spectral lines including field-dependent atomic physics and the ion quadrupole interaction}, volume={48}, ISSN={1063-651X, 1095-3787}, DOI={10.1103/PhysRevE.48.3901}, number={5}, journal={Physical Review E}, author={Kilcrease, D. P. and Mancini, R. C. and Hooper, C. F.}, year={1993}, month=nov, pages={3901–3913}, language={en} }

@ARTICLE{Ferri14,
       author = {{Ferri}, Sandrine and {Calisti}, Annette and {Moss{\'e}}, Caroline and {Rosato}, Jo{\"e}l and {Talin}, Bernard and {Alexiou}, Spiros and {Gigosos}, Marco and {Gonz{\'a}lez}, Manuel and {Gonz{\'a}lez-Herrero}, Diego and {Lara}, Natividad and {Gomez}, Thomas and {Iglesias}, Carlos and {Lorenzen}, Sonja and {Mancini}, Roberto and {Stambulchik}, Evgeny},
        title = "{Ion Dynamics Effect on Stark-Broadened Line Shapes: A Cross-Comparison of Various Models}",
      journal = {Atoms},
         year = 2014,
        month = jul,
       volume = {2},
       number = {3},
        pages = {299-318},
          doi = {10.3390/atoms2030299},
       adsurl = {https://ui.adsabs.harvard.edu/abs/2014Atoms...2..299F}
}

@article{Ferri21,
    author = {Ferri, Sandrine and Peyrusse, Olivier and Calisti, Annette},
    title = "{Stark–Zeeman line-shape model for multi-electron radiators in hot dense plasmas subjected to large magnetic fields}",
    journal = {Matter and Radiation at Extremes},
    volume = {7},
    number = {1},
    pages = {015901},
    year = {2021},
    month = {12},
    issn = {2468-2047},
    doi = {10.1063/5.0058552},
    url = {https://doi.org/10.1063/5.0058552},
    eprint = {https://pubs.aip.org/aip/mre/article-pdf/doi/10.1063/5.0058552/16615884/015901\_1\_online.pdf},
}

@ARTICLE{Santos12,
       author = {{Santos}, M.~G. and {Kepler}, S.~O.},
        title = "{Theoretical study of the line profiles of the hydrogen perturbed by collisions with protons}",
      journal = {\mnras},
         year = 2012,
        month = jun,
       volume = {423},
       number = {1},
        pages = {68-79},
          doi = {10.1111/j.1365-2966.2012.20631.x},
archivePrefix = {arXiv},
       eprint = {1204.4769},
 primaryClass = {astro-ph.SR},
       adsurl = {https://ui.adsabs.harvard.edu/abs/2012MNRAS.423...68S}
}

@ARTICLE{Falcon13,
       author = {{Falcon}, Ross E. and {Rochau}, G.~A. and {Bailey}, J.~E. and {Ellis}, J.~L. and {Carlson}, A.~L. and {Gomez}, T.~A. and {Montgomery}, M.~H. and {Winget}, D.~E. and {Chen}, E.~Y. and {Gomez}, M.~R. and {Nash}, T.~J.},
        title = "{An experimental platform for creating white dwarf photospheres in the laboratory}",
      journal = {High Energy Density Physics},
         year = 2013,
        month = mar,
       volume = {9},
       number = {1},
        pages = {82-90},
          doi = {10.1016/j.hedp.2012.10.005},
archivePrefix = {arXiv},
       eprint = {1210.7197},
 primaryClass = {astro-ph.SR},
       adsurl = {https://ui.adsabs.harvard.edu/abs/2013HEDP....9...82F}
}

@ARTICLE{Baranger58,
       author = {{Baranger}, Michel},
        title = "{Problem of Overlapping Lines in the Theory of Pressure Broadening}",
      journal = {Physical Review},
         year = 1958,
        month = jul,
       volume = {111},
       number = {2},
        pages = {494-504},
          doi = {10.1103/PhysRev.111.494},
       adsurl = {https://ui.adsabs.harvard.edu/abs/1958PhRv..111..494B}
}

@ARTICLE{Fano63,
       author = {{Fano}, U.},
        title = "{Pressure Broadening as a Prototype of Relaxation}",
      journal = {Physical Review},
         year = 1963,
        month = jul,
       volume = {131},
       number = {1},
        pages = {259-268},
          doi = {10.1103/PhysRev.131.259},
       adsurl = {https://ui.adsabs.harvard.edu/abs/1963PhRv..131..259F}
}

@article{Rosato20, 
title={Quantifying the statistical noise in computer simulations of Stark broadening}, volume={249}, ISSN={00224073}, DOI={10.1016/j.jqsrt.2020.107002}, journal={Journal of Quantitative Spectroscopy and Radiative Transfer}, author={Rosato, J. and Marandet, Y. and Stamm, R.}, year={2020}, month=jul, pages={107002}, language={en} 
}

@BOOK{Griem74,
       author = {{Griem}, H.~R.},
        title = "{Spectral line broadening by plasmas}",
         year = 1974,
       adsurl = {https://ui.adsabs.harvard.edu/abs/1974slbp.book.....G}
}

@ARTICLE{Seidel82,
       author = {{Seidel}, J. and {Stamm}, R.},
        title = "{Effects of radiator motion on plasma-broadened hydrogen lyman-{\ensuremath{\beta}}}",
      journal = {\jqsrt},
         year = 1982,
        month = may,
       volume = {27},
       number = {5},
        pages = {499-503},
          doi = {10.1016/0022-4073(82)90102-9},
       adsurl = {https://ui.adsabs.harvard.edu/abs/1982JQSRT..27..499S}
}

@ARTICLE{Wiese72,
       author = {{Wiese}, W.~L. and {Kelleher}, D.~E. and {Paquette}, D.~R.},
        title = "{Detailed Study of the Stark Broadening of Balmer Lines in a High-Density Plasma}",
      journal = {\pra},
         year = 1972,
        month = sep,
       volume = {6},
       number = {3},
        pages = {1132-1153},
          doi = {10.1103/PhysRevA.6.1132},
       adsurl = {https://ui.adsabs.harvard.edu/abs/1972PhRvA...6.1132W}
}

@ARTICLE{Bergeron19,
       author = {{Bergeron}, P. and {Dufour}, P. and {Fontaine}, G. and {Coutu}, S. and {Blouin}, S. and {Genest-Beaulieu}, C. and {B{\'e}dard}, A. and {Rolland}, B.},
        title = "{On the Measurement of Fundamental Parameters of White Dwarfs in the Gaia Era}",
      journal = {\apj},
         year = 2019,
        month = may,
       volume = {876},
       number = {1},
          eid = {67},
        pages = {67},
          doi = {10.3847/1538-4357/ab153a},
archivePrefix = {arXiv},
       eprint = {1904.02022},
 primaryClass = {astro-ph.SR},
       adsurl = {https://ui.adsabs.harvard.edu/abs/2019ApJ...876...67B}
}

@ARTICLE{Saumon22,
       author = {{Saumon}, Didier and {Blouin}, Simon and {Tremblay}, Pier-Emmanuel},
        title = "{Current challenges in the physics of white dwarf stars}",
      journal = {\physrep},
         year = 2022,
        month = nov,
       volume = {988},
        pages = {1-63},
          doi = {10.1016/j.physrep.2022.09.001},
archivePrefix = {arXiv},
       eprint = {2209.02846},
 primaryClass = {astro-ph.SR},
       adsurl = {https://ui.adsabs.harvard.edu/abs/2022PhR...988....1S}
}

@ARTICLE{Schaeuble22,
       author = {{Schaeuble}, M. -A. and {Nagayama}, T. and {Bailey}, J.~E. and {Gigosos}, M.~A. and {Florido}, R. and {Blouin}, S. and {Gomez}, T.~A. and {Dunlap}, B. and {Montgomery}, M.~H. and {Winget}, D.~E.},
        title = "{Measuring He I Stark Line Shapes in the Laboratory to Examine Differences in Photometric and Spectroscopic DB White Dwarf Masses}",
      journal = {\apj},
         year = 2022,
        month = dec,
       volume = {940},
       number = {2},
          eid = {181},
        pages = {181},
          doi = {10.3847/1538-4357/ac9df5},
       adsurl = {https://ui.adsabs.harvard.edu/abs/2022ApJ...940..181S}
}

@INPROCEEDINGS{LeboucherDalimier99,
       author = {{Leboucher-Dalimier}, E. and {Sauvan}, P. and {Angelo}, P. and {Derfoul}, H. and {Ceccotti}, T. and {Gauthier}, P. and {Poqu{\'e}russe}, A. and {Calisti}, A. and {Talin}, B.},
        title = "{Alternative treatment of line broadening in dense and hot plasmas}",
    booktitle = {Spectral Line Shapes},
         year = 1999,
       editor = {{Herman}, Roger M.},
       series = {American Institute of Physics Conference Series},
       volume = {467},
        month = apr,
    publisher = {AIP},
        pages = {64-76},
          doi = {10.1063/1.58349},
       adsurl = {https://ui.adsabs.harvard.edu/abs/1999AIPC..467...64L}
}

@ARTICLE{Allard98_2,
       author = {{Allard}, N.~F. and {Kielkopf}, J. and {Feautrier}, N.},
        title = "{Satellites on the Lyman beta line of atomic hydrogen due to H-H(+) collisions}",
      journal = {\aap},
         year = 1998,
        month = feb,
       volume = {330},
        pages = {782-790},
       adsurl = {https://ui.adsabs.harvard.edu/abs/1998A&A...330..782A}
}

@ARTICLE{Allard2000,
       author = {{Allard}, N.~F. and {Kielkopf}, J. and {Drira}, I. and {Schmelcher}, P.},
        title = "{A collision-induced satellite in the Lyman   profile due to H-H collisions <SBT>Lyman satellites</SBT>}",
      journal = {European Physical Journal D},
         year = 2000,
        month = jan,
       volume = {12},
       number = {2},
        pages = {263-268},
          doi = {10.1007/s100530070021},
archivePrefix = {arXiv},
       eprint = {physics/0006028},
 primaryClass = {physics.atom-ph},
       adsurl = {https://ui.adsabs.harvard.edu/abs/2000EPJD...12..263A}
}

@BOOK{VanHorn15,
       author = {{Van Horn}, Hugh M.},
        title = "{Unlocking the Secrets of White Dwarf Stars}",
         year = 2015,
publisher = "{Cham:Springer}",
          doi = {10.1007/978-3-319-09369-7},
       adsurl = {https://ui-adsabs-harvard-edu.ezproxy.lib.utexas.edu/abs/2015uswd.book.....V}
}

@ARTICLE{Winget87,
       author = {{Winget}, D.~E. and {Hansen}, C.~J. and {Liebert}, James and {van Horn}, H.~M. and {Fontaine}, G. and {Nather}, R.~E. and {Kepler}, S.~O. and {Lamb}, D.~Q.},
        title = "{An Independent Method for Determining the Age of the Universe}",
      journal = {\apjl},
         year = 1987,
        month = apr,
       volume = {315},
        pages = {L77},
          doi = {10.1086/184864},
       adsurl = {https://ui.adsabs.harvard.edu/abs/1987ApJ...315L..77W}
}

@ARTICLE{Gordon22,
       author = {{Gordon}, Karl D. and {Bohlin}, Ralph and {Sloan}, G.~C. and {Rieke}, George and {Volk}, Kevin and {Boyer}, Martha and {Muzerolle}, James and {Schlawin}, Everett and {Deustua}, Susana E. and {Hines}, Dean C. and {Kraemer}, Kathleen E. and {Mullally}, Susan E. and {Su}, Kate Y.~L.},
        title = "{The James Webb Space Telescope Absolute Flux Calibration. I. Program Design and Calibrator Stars}",
      journal = {\aj},
         year = 2022,
        month = jun,
       volume = {163},
       number = {6},
          eid = {267},
        pages = {267},
          doi = {10.3847/1538-3881/ac66dc},
archivePrefix = {arXiv},
       eprint = {2204.06500},
 primaryClass = {astro-ph.IM},
       adsurl = {https://ui.adsabs.harvard.edu/abs/2022AJ....163..267G}
}

@ARTICLE{Zuckerman07,
       author = {{Zuckerman}, B. and {Koester}, D. and {Melis}, C. and {Hansen}, Brad M. and {Jura}, M.},
        title = "{The Chemical Composition of an Extrasolar Minor Planet}",
      journal = {\apj},
         year = 2007,
        month = dec,
       volume = {671},
       number = {1},
        pages = {872-877},
          doi = {10.1086/522223},
archivePrefix = {arXiv},
       eprint = {0708.0198},
 primaryClass = {astro-ph},
       adsurl = {https://ui.adsabs.harvard.edu/abs/2007ApJ...671..872Z}
}

@ARTICLE{Farihi13,
       author = {{Farihi}, J. and {G{\"a}nsicke}, B.~T. and {Koester}, D.},
        title = "{Evidence for Water in the Rocky Debris of a Disrupted Extrasolar Minor Planet}",
      journal = {Science},
         year = 2013,
        month = oct,
       volume = {342},
       number = {6155},
        pages = {218-220},
          doi = {10.1126/science.1239447},
archivePrefix = {arXiv},
       eprint = {1310.3269},
 primaryClass = {astro-ph.EP},
       adsurl = {https://ui.adsabs.harvard.edu/abs/2013Sci...342..218F}
}

@ARTICLE{Falcon15,
       author = {{Falcon}, Ross E. and {Rochau}, G.~A. and {Bailey}, J.~E. and {Gomez}, T.~A. and {Montgomery}, M.~H. and {Winget}, D.~E. and {Nagayama}, T.},
        title = "{Laboratory Measurements of White Dwarf Photospheric Spectral Lines: H{\ensuremath{\beta}}}",
      journal = {\apj},
         year = 2015,
        month = jun,
       volume = {806},
       number = {2},
          eid = {214},
        pages = {214},
          doi = {10.1088/0004-637X/806/2/214},
archivePrefix = {arXiv},
       eprint = {1505.03801},
 primaryClass = {astro-ph.SR},
       adsurl = {https://ui.adsabs.harvard.edu/abs/2015ApJ...806..214F}
}

@ARTICLE{Hubeny21,
       author = {{Hubeny}, Ivan and {Allende Prieto}, Carlos and {Osorio}, Yeisson and {Lanz}, Thierry},
        title = "{TLUSTY and SYNSPEC Users's Guide IV: Upgraded Versions 208 and 54}",
      journal = {arXiv e-prints},
         year = 2021,
        month = apr,
          eid = {arXiv:2104.02829},
        pages = {arXiv:2104.02829},
          doi = {10.48550/arXiv.2104.02829},
archivePrefix = {arXiv},
       eprint = {2104.02829},
 primaryClass = {astro-ph.SR},
       adsurl = {https://ui-adsabs-harvard-edu.ezproxy.lib.utexas.edu/abs/2021arXiv210402829H}
}

@ARTICLE{Gomez2025,
       author = {{Gomez}, Thomas A. and {Zammit}, Mark C. and {White}, Jackson R. and {Stambulchik}, Evgeny and {Hubeny}, Ivan and {Bray}, Igor and {Fontes}, Christopher J. and {Montgomery}, Michael H. and {Dunlap}, Bart H. and {Winget}, Donald E.},
        title = "{Increased Ly{\ensuremath{\alpha}} Opacity in White Dwarf Photospheres from Transient H$^{‑}$ Resonances}",
      journal = {\apj},
         year = 2025,
        month = jun,
       volume = {986},
       number = {1},
          eid = {52},
        pages = {52},
          doi = {10.3847/1538-4357/adc9ae},
       adsurl = {https://ui.adsabs.harvard.edu/abs/2025ApJ...986...52G}
}

@misc{KoesterPrivate,
  author = "Koester, D.",
  year = 2026,
  howpublished = "private communication"
}

@ARTICLE{Moos2000,
       author = {{Moos}, H.~W. and {Cash}, W.~C. and {Cowie}, L.~L. and {Davidsen}, A.~F. and {Dupree}, A.~K. and {Feldman}, P.~D. and {Friedman}, S.~D. and {Green}, J.~C. and {Green}, R.~F. and {Gry}, C. and {Hutchings}, J.~B. and {Jenkins}, E.~B. and {Linsky}, J.~L. and {Malina}, R.~F. and {Michalitsianos}, A.~G. and {Savage}, B.~D. and {Shull}, J.~M. and {Siegmund}, O.~H.~W. and {Snow}, T.~P. and {Sonneborn}, G. and {Vidal-Madjar}, A. and {Willis}, A.~J. and {Woodgate}, B.~E. and {York}, D.~G. and {Ake}, T.~B. and {Andersson}, B. -G. and {Andrews}, J.~P. and {Barkhouser}, R.~H. and {Bianchi}, L. and {Blair}, W.~P. and {Brownsberger}, K.~R. and {Cha}, A.~N. and {Chayer}, P. and {Conard}, S.~J. and {Fullerton}, A.~W. and {Gaines}, G.~A. and {Grange}, R. and {Gummin}, M.~A. and {Hebrard}, G. and {Kriss}, G.~A. and {Kruk}, J.~W. and {Mark}, D. and {McCarthy}, D.~K. and {Morbey}, C.~L. and {Murowinski}, R. and {Murphy}, E.~M. and {Oegerle}, W.~R. and {Ohl}, R.~G. and {Oliveira}, C. and {Osterman}, S.~N. and {Sahnow}, D.~J. and {Saisse}, M. and {Sembach}, K.~R. and {Weaver}, H.~A. and {Welsh}, B.~Y. and {Wilkinson}, E. and {Zheng}, W.},
        title = "{Overview of the Far Ultraviolet Spectroscopic Explorer Mission}",
      journal = {\apjl},
         year = 2000,
        month = jul,
       volume = {538},
       number = {1},
        pages = {L1-L6},
          doi = {10.1086/312795},
archivePrefix = {arXiv},
       eprint = {astro-ph/0005529},
 primaryClass = {astro-ph},
       adsurl = {https://ui.adsabs.harvard.edu/abs/2000ApJ...538L...1M}
}

@ARTICLE{Sahnow2000,
       author = {{Sahnow}, D.~J. and {Moos}, H.~W. and {Ake}, T.~B. and {Andersen}, J. and {Andersson}, B. -G. and {Andre}, M. and {Artis}, D. and {Berman}, A.~F. and {Blair}, W.~P. and {Brownsberger}, K.~R. and {Calvani}, H.~M. and {Chayer}, P. and {Conard}, S.~J. and {Feldman}, P.~D. and {Friedman}, S.~D. and {Fullerton}, A.~W. and {Gaines}, G.~A. and {Gawne}, W.~C. and {Green}, J.~C. and {Gummin}, M.~A. and {Jennings}, T.~B. and {Joyce}, J.~B. and {Kaiser}, M.~E. and {Kruk}, J.~W. and {Lindler}, D.~J. and {Massa}, D. and {Murphy}, E.~M. and {Oegerle}, W.~R. and {Ohl}, R.~G. and {Roberts}, B.~A. and {Romelfanger}, M.~L. and {Roth}, K.~C. and {Sankrit}, R. and {Sembach}, K.~R. and {Shelton}, R.~L. and {Siegmund}, O.~H.~W. and {Silva}, C.~J. and {Sonneborn}, G. and {Vaclavik}, S.~R. and {Weaver}, H.~A. and {Wilkinson}, E.},
        title = "{On-Orbit Performance of the Far Ultraviolet Spectroscopic Explorer Satellite}",
      journal = {\apjl},
         year = 2000,
        month = jul,
       volume = {538},
       number = {1},
        pages = {L7-L11},
          doi = {10.1086/312794},
archivePrefix = {arXiv},
       eprint = {astro-ph/0005531},
 primaryClass = {astro-ph},
       adsurl = {https://ui.adsabs.harvard.edu/abs/2000ApJ...538L...7S}
}

@ARTICLE{Hebrard2003,
       author = {{H{\'e}brard}, G. and {Allard}, N.~F. and {Kielkopf}, J.~F. and {Chayer}, P. and {Dupuis}, J. and {Kruk}, J.~W. and {Hubeny}, I.},
        title = "{Modeling of the Lyman gamma  satellites in FUSE spectra of DA white dwarfs}",
      journal = {\aap},
         year = 2003,
        month = jul,
       volume = {405},
        pages = {1153-1156},
          doi = {10.1051/0004-6361:20030715},
archivePrefix = {arXiv},
       eprint = {astro-ph/0305356},
 primaryClass = {astro-ph},
       adsurl = {https://ui.adsabs.harvard.edu/abs/2003A&A...405.1153H}
}

@ARTICLE{Giammichele2012,
       author = {{Giammichele}, N. and {Bergeron}, P. and {Dufour}, P.},
        title = "{Know Your Neighborhood: A Detailed Model Atmosphere Analysis of Nearby White Dwarfs}",
      journal = {\apjs},
         year = 2012,
        month = apr,
       volume = {199},
       number = {2},
          eid = {29},
        pages = {29},
          doi = {10.1088/0067-0049/199/2/29},
archivePrefix = {arXiv},
       eprint = {1202.5581},
 primaryClass = {astro-ph.SR},
       adsurl = {https://ui.adsabs.harvard.edu/abs/2012ApJS..199...29G}
}

@ARTICLE{Allard2022,
       author = {{Allard}, N.~F. and {Spiegelman}, F. and {Kielkopf}, J.~F. and {Bourdreux}, S.},
        title = "{Collisional effects in the blue wing of the Balmer-{\ensuremath{\alpha}} line}",
      journal = {\aap},
         year = 2022,
        month = jan,
       volume = {657},
          eid = {A121},
        pages = {A121},
          doi = {10.1051/0004-6361/202141461},
archivePrefix = {arXiv},
       eprint = {2201.00878},
 primaryClass = {astro-ph.SR},
       adsurl = {https://ui.adsabs.harvard.edu/abs/2022A&A...657A.121A}
}

@ARTICLE{Allard2004,
       author = {{Allard}, N.~F. and {Kielkopf}, J.~F. and {Loeillet}, B.},
        title = "{Temperature dependence of the Lyman {\ensuremath{\alpha}} line wings in cool hydrogen-rich white dwarf atmospheres. Application to ZZ Ceti white dwarf spectra}",
      journal = {\aap},
         year = 2004,
        month = sep,
       volume = {424},
        pages = {347-354},
          doi = {10.1051/0004-6361:20040427},
       adsurl = {https://ui.adsabs.harvard.edu/abs/2004A&A...424..347A}
}

@ARTICLE{Waltz1984,
       author = {{Woltz}, L.~A. and {Hooper}, Jr., C.~F.},
        title = "{Full Coulomb calculation of Stark broadening in laser-produced plasmas}",
      journal = {\pra},
         year = 1984,
        month = jul,
       volume = {30},
       number = {1},
        pages = {468-473},
          doi = {10.1103/PhysRevA.30.468},
       adsurl = {https://ui.adsabs.harvard.edu/abs/1984PhRvA..30..468W}
}

@ARTICLE{Demura2014,
       author = {{Demura}, Alexander V. and {Stambulchik}, Evgeny},
        title = "{Spectral-Kinetic Coupling and Effect of Microfield Rotation on Stark Broadening in Plasmas}",
      journal = {Atoms},
         year = 2014,
        month = jul,
       volume = {2},
       number = {3},
        pages = {334-356},
          doi = {10.3390/atoms2030334},
       adsurl = {https://ui.adsabs.harvard.edu/abs/2014Atoms...2..334D}
}

@ARTICLE{Vanderbosch2020,
       author = {{Vanderbosch}, Z. and {Hermes}, J.~J. and {Dennihy}, E. and {Dunlap}, B.~H. and {Izquierdo}, P. and {Tremblay}, P.-E. and {Cho}, P.~B. and {G{\"a}nsicke}, B.~T. and {Toloza}, O. and {Bell}, K.~J. and {Montgomery}, M.~H. and {Winget}, D.~E.},
        title = "{A White Dwarf with Transiting Circumstellar Material Far outside the Roche Limit}",
      journal = {\apj},
         year = 2020,
        month = jul,
       volume = {897},
       number = {2},
          eid = {171},
        pages = {171},
          doi = {10.3847/1538-4357/ab9649},
archivePrefix = {arXiv},
       eprint = {1908.09839},
 primaryClass = {astro-ph.SR},
       adsurl = {https://ui.adsabs.harvard.edu/abs/2020ApJ...897..171V}
}

@ARTICLE{arseneau25,
       author = {{Arseneau}, Stefan M. and {Hermes}, J.~J. and {Zakamska}, Nadia L. and {El-Badry}, Kareem and {Crumpler}, Nicole R. and {Chandra}, Vedant and {Adamane Pallathadka}, Gautham and {Badenes}, Carles and {G{\"a}nsicke}, Boris T. and {Fusillo}, Nicola Gentile},
        title = "{Resolution-corrected White Dwarf Gravitational Redshifts Validate Fifth-generation Sloan Digital Sky Survey Wavelength Calibration and Enable Accurate Mass{\textendash}Radius Tests}",
      journal = {\apj},
         year = 2025,
        month = oct,
       volume = {991},
       number = {2},
          eid = {190},
        pages = {190},
          doi = {10.3847/1538-4357/adf8dd},
archivePrefix = {arXiv},
       eprint = {2508.04775},
 primaryClass = {astro-ph.SR},
       adsurl = {https://ui.adsabs.harvard.edu/abs/2025ApJ...991..190A}
}

@ARTICLE{holweger94,
       author = {{Holweger}, H. and {Koester}, D. and {Allard}, N.~F.},
        title = "{Identification of the 1600A feature in Lambda Bootis stars}",
      journal = {\aap},
         year = 1994,
        month = oct,
       volume = {290},
        pages = {L21-L24},
       adsurl = {https://ui.adsabs.harvard.edu/abs/1994A&A...290L..21H}
}

@ARTICLE{Gigosos18,
       author = {{Gigosos}, M.~A. and {Gonz{\'a}lez-Herrero}, D. and {Lara}, N. and {Florido}, R. and {Calisti}, A. and {Ferri}, S. and {Talin}, B.},
        title = "{Classical molecular dynamics simulations of hydrogen plasmas and development of an analytical statistical model for computational validity assessment}",
      journal = {\pre},
         year = 2018,
        month = sep,
       volume = {98},
       number = {3},
          eid = {033307},
        pages = {033307},
          doi = {10.1103/PhysRevE.98.033307},
archivePrefix = {arXiv},
       eprint = {2402.03401},
 primaryClass = {physics.plasm-ph},
       adsurl = {https://ui.adsabs.harvard.edu/abs/2018PhRvE..98c3307G}
}

@software{Hubeny2011,
       author = {{Hubeny}, Ivan and {Lanz}, Thierry},
        title = "{Synspec: General Spectrum Synthesis Program}",
 howpublished = {Astrophysics Source Code Library, record ascl:1109.022},
         year = 2011,
        month = sep,
          eid = {ascl:1109.022},
archivePrefix = {ascl},
       eprint = {1109.022},
       adsurl = {https://ui.adsabs.harvard.edu/abs/2011ascl.soft09022H}
}

@ARTICLE{Vidal1970,
       author = {{Vidal}, C.~R. and {Cooper}, J. and {Smith}, E.~W.},
        title = "{Hydrogen Stark broadening calculations with the unified classical path theory}",
      journal = {\jqsrt},
         year = 1970,
        month = jan,
       volume = {10},
       number = {9},
        pages = {1011-1063},
          doi = {10.1016/0022-4073(70)90121-4},
       adsurl = {https://ui.adsabs.harvard.edu/abs/1970JQSRT..10.1011V}
}

@article{Zammit2013,
  title = {Calculations of electron scattering from H${{}_{2}}^{+}$},
  author = {Zammit, Mark C. and Fursa, Dmitry V. and Bray, Igor},
  journal = {Phys. Rev. A},
  volume = {88},
  issue = {6},
  pages = {062709},
  numpages = {4},
  year = {2013},
  month = {Dec},
  publisher = {American Physical Society},
  doi = {10.1103/PhysRevA.88.062709},
  url = {https://link.aps.org/doi/10.1103/PhysRevA.88.062709}
}

@article{Zammit2014,
  title = {Electron scattering from the molecular hydrogen ion and its isotopologues},
  author = {Zammit, Mark C. and Fursa, Dmitry V. and Bray, Igor},
  journal = {Phys. Rev. A},
  volume = {90},
  issue = {2},
  pages = {022711},
  numpages = {15},
  year = {2014},
  month = {Aug},
  publisher = {American Physical Society},
  doi = {10.1103/PhysRevA.90.022711},
  url = {https://link.aps.org/doi/10.1103/PhysRevA.90.022711}
}

@ARTICLE{Tremblay2009,
       author = {{Tremblay}, P.-E. and {Bergeron}, P.},
        title = "{Spectroscopic Analysis of DA White Dwarfs: Stark Broadening of Hydrogen Lines Including Nonideal Effects}",
      journal = {\apj},
         year = 2009,
        month = may,
       volume = {696},
       number = {2},
        pages = {1755-1770},
          doi = {10.1088/0004-637X/696/2/1755},
archivePrefix = {arXiv},
       eprint = {0902.4182},
 primaryClass = {astro-ph.SR},
       adsurl = {https://ui.adsabs.harvard.edu/abs/2009ApJ...696.1755T}
}

@ARTICLE{Ramaker72,
       author = {{Ramaker}, D.~E. and {Peek}, J.~M.},
        title = "{$^{2}$H$_{2}$$^{+}$ dipole strengths by asymptotic techniques}",
      journal = {Journal of Physics B Atomic Molecular Physics},
         year = 1972,
        month = dec,
       volume = {5},
       number = {12},
        pages = {2175-2181},
          doi = {10.1088/0022-3700/5/12/011},
       adsurl = {https://ui.adsabs.harvard.edu/abs/1972JPhB....5.2175R}
}

@ARTICLE{Madsen71,
       author = {{Madsen}, M.~M. and {Peek}, J.~M.},
        title = "{Eigenparameters for the Lowest Twenty Electronic States of the Hydrogen Molecule Ion}",
      journal = {Atomic Data},
         year = 1971,
        month = jan,
       volume = {2},
        pages = {171},
          doi = {10.1016/S0092-640X(70)80008-0},
       adsurl = {https://ui.adsabs.harvard.edu/abs/1971AD......2..171M}
}

@ARTICLE{Tremblay2010,
       author = {{Tremblay}, P.-E. and {Bergeron}, P. and {Kalirai}, J.~S. and {Gianninas}, A.},
        title = "{New Insights into the Problem of the Surface Gravity Distribution of Cool DA White Dwarfs}",
      journal = {\apj},
         year = 2010,
        month = apr,
       volume = {712},
       number = {2},
        pages = {1345-1358},
          doi = {10.1088/0004-637X/712/2/1345},
archivePrefix = {arXiv},
       eprint = {1002.3585},
 primaryClass = {astro-ph.SR},
       adsurl = {https://ui.adsabs.harvard.edu/abs/2010ApJ...712.1345T}
}

@ARTICLE{Gomez16,
       author = {{Gomez}, T.~A. and {Nagayama}, T. and {Kilcrease}, D.~P. and {Montgomery}, M.~H. and {Winget}, D.~E.},
        title = "{Effect of higher-order multipole moments on the Stark line shape}",
      journal = {\pra},
         year = 2016,
        month = aug,
       volume = {94},
       number = {2},
          eid = {022501},
        pages = {022501},
          doi = {10.1103/PhysRevA.94.022501},
       adsurl = {https://ui.adsabs.harvard.edu/abs/2016PhRvA..94b2501G}
}

@MISC{Holberg01,
author = {{Holberg}, Jay B.},
title = "{A FUSE Search for Nitrogen and Carbon in Hot White Dwarfs}",
howpublished = {FUSE Proposal ID B119},
year = 2001,
month = jan,
adsurl = {https://ui.adsabs.harvard.edu/abs/2001fuse.prop.B119H}
}

@BOOK{Cowan1981,
       author = {{Cowan}, Robert D.},
        title = "{The theory of atomic structure and spectra}",
         year = 1981,
       adsurl = {https://ui.adsabs.harvard.edu/abs/1981tass.book.....C}
}

@ARTICLE{Blouin2019,
       author = {{Blouin}, S. and {Allard}, N.~F. and {Leininger}, T. and {Gad{\'e}a}, F.~X. and {Dufour}, P.},
        title = "{Line Profiles of the Calcium I Resonance Line in Cool Metal-polluted White Dwarfs}",
      journal = {\apj},
         year = 2019,
        month = apr,
       volume = {875},
       number = {2},
          eid = {137},
        pages = {137},
          doi = {10.3847/1538-4357/ab1266},
archivePrefix = {arXiv},
       eprint = {1903.08503},
 primaryClass = {astro-ph.SR},
       adsurl = {https://ui.adsabs.harvard.edu/abs/2019ApJ...875..137B}
}

@ARTICLE{Allard2018,
       author = {{Allard}, N.~F. and {Kielkopf}, J.~F. and {Blouin}, S. and {Dufour}, P. and {Gad{\'e}a}, F.~X. and {Leininger}, T. and {Guillon}, G.},
        title = "{Line shapes of the magnesium resonance lines in cool DZ white dwarf atmospheres}",
      journal = {\aap},
         year = 2018,
        month = nov,
       volume = {619},
          eid = {A152},
        pages = {A152},
          doi = {10.1051/0004-6361/201834067},
archivePrefix = {arXiv},
       eprint = {1809.04531},
 primaryClass = {astro-ph.SR},
       adsurl = {https://ui.adsabs.harvard.edu/abs/2018A&A...619A.152A}
}
\bibliographystyle{aasjournalv7}



\end{document}